\documentclass{aa}  
\usepackage{graphicx}
\usepackage{natbib}
\usepackage{ulem}
\usepackage{txfonts}
\begin{document}
   \title{PCA detection and denoising of Zeeman signatures in stellar polarised spectra}

   \subtitle{}

   \author{M. J. Mart\' inez Gonz\'alez\inst{1}
          \and
          A. Asensio Ramos\inst{2}
	  \and
	  T. A. Carroll\inst{3}
	  \and
	  M. Kopf\inst{3}
	  \and 
	  J. C. Ram\' irez V\'elez\inst{4}
	  \and
	  M. Semel\inst{4}}

   \offprints{M. J. Mart\' inez Gonz\'alez}

   \institute{LERMA, Observatoire de Paris-Meudon, 5 place de Jules Janssen,
92195, Meudon, France\\
              \email{Marian.Martinez@obspm.fr}
	      \and 
	      Instituto de Astrof\' isica de Canarias, V\' ia L\'actea S/N,
38200, La Laguna, Spain\\
	      \email{aasensio@iac.es}
	      \and
              Astrophysikalisches Institut Potsdam, An der Sternwarte 16, 14482 Potsdam, Germany\\
	      \email{tcarroll@aip.de}
              \and
	      LESIA, Observatoire de Paris-Meudon, 5 place de Jules Janssen,
92195, Meudon, France\\
              \email{Julio.Ramirez@obspm.fr, Meir.Semel@obspm.fr}}

   \date{Received ; Accepted }

 
  \abstract
   {}
   {Our main objective is to develop a denoising strategy to increase the signal to noise ratio of
individual spectral lines of stellar spectropolarimetric observations.}
   {We use a multivariate statistics technique called Principal Component Analysis. The cross-product matrix of
the observations is diagonalized to obtain the eigenvectors in which the original observations can be
developed. This basis is such that the first eigenvectors contain the greatest variance. Assuming that
the noise is uncorrelated a denoising is possible by reconstructing the data with a truncated basis.
We propose a method to identify the number of eigenvectors for an efficient noise filtering.}
   {Numerical simulations are used to demonstrate that an important increase of the signal to noise
ratio per spectral line is possible using PCA denoising techniques. It can be also applied for detection
of magnetic fields in stellar atmospheres. We analyze the relation between PCA and commonly used
well-known techniques like line addition and least-squares deconvolution. Moreover, PCA is very robust 
and easy to compute. }
   {}

   \keywords{Polarization --- Stars: magnetic fields --- Methods: numerical}

   \maketitle
%

\section{Introduction}

The light coming from most of the astrophysical scenarios is polarised. This
radiation is usually described in terms of the Stokes parameters: the total
intensity, $I$, the linear polarisation given by Stokes $Q$ and $U$, and the
circular polarisation given by Stokes $V$. The degree of polarisation is
accounted for $\sqrt{Q^2+U^2+V^2}$ which, in most cases, is very small compared
to the total intensity. Several physical mechanisms related to the breaking of
the spherical symmetry induce the generation of polarised radiation: scattering
processes, the presence of magnetic fields through the Zeeman or Hanle effects,
etc. For example, the presence of strong magnetic fields in solar or stellar
spots produces a large degree of polarisation which, in some cases, can reach
some tens percent. It decreases to $\sim 0.01-0.1$ \% in the less magnetically
active areas of solar or stellar surfaces.

The investigation of the magnetic field in stellar atmospheres is restrained by
the low expected and observed polarisation signals \cite[e.g.,][]{donati97}. In most cases the
expected degree of polarisation is of the order or even below the noise level.
This is specially critical when analysing the spectra of cool stars 
\citep[][and references therein]{petit07},
although this problem is already present for active stars. The most
natural procedure to increase the signal to noise ratio ($S/N$) is to increase the
exposure time. However, it is limited by the rotation period of the star. A
radically new solution to this problem was presented by \cite{semel96} \citep[see also][]{semel89}. They suggested to
use multiline observations of the same star and combine the information of all
of them to increase the sensitivity of the polarimetric observations. In the
last decades, these ideas have been made possible thanks to the synergy between
instrumental and theoretical advances. On the one hand, we have witnessed the
development of very sensitive polarimeters attached to 
cross-dispersed Echelle spectrographs that produce data of very good scientific
quality. Some examples of very successful instruments are
ESPaDOnS\footnote{Echelle SpectroPolarimetric Device for the Observation of
Stars} and NARVAL, both based on the concept of SEMPOL\footnote{SEMel
POLarimeter}. On the other hand, line addition techniques have permitted to take
advantage of these large spectral range observations. An inflection point was
the presentation of the Least-Squares Deconvolution (LSD) technique that allowed
to detect polarisation signals in a broad variety of stars \citep{donati97,donati07}. The LSD
technique is an improvement over the brute force line addition approach
developped by \cite{semel96}. 

Although line addition techniques are very successful on the detection of the
polarisation signatures in noisy spectra, they are based on very rough
approximations. Another weak point is that the final polarisation signature
obtained after applying these techniques is difficult to interpret. It cannot be
associated with a standard spectral line and any analysis based on the theory of
polarised radiative transfer of spectral lines cannot be directly applied. For
this reason, it would be desirable not to work with ``mean'' profiles but to
take advantage of multiline observations to increase the $S/N$ of individual
spectral lines. This would make much easier to interpret the results because one
would deal with standard spectral lines. 

Following this idea, we present in this paper a denoising technique based on a
multivariate statistics method called Principal Component Analysis (PCA). A first application of this technique on real observational
data has been used by \cite{carroll08} for Zeeman-Doppler imaging of late-type stars. A technique based on PCA has been already proposed for increasing the 
signal to noise ratio of stellar polarised spectra by \cite{semel06}. This technique uses a 
data base of synthetic stellar spectra to construct the so-called Multi Zeeman Signatures, 
profiles that result from the cross-correlation of the observed spectrum and the principal components of 
the synthetic data base. In this paper, we propose another different approach using the same 
statistical technique. It allows us to efficiently denoise stellar
polarised spectra so that an increase of the $S/N$ per individual spectral line is obtained
using efficiently the information encoded in the whole observed spectrum. Consequently, since we use 
the observed stellar spectrum itself, our method is model-independent. Moreover, we denoise the 
whole spectrum keeping the information carried by the spectral 
lines, making them susceptible of more sophisticated radiative transfer analysis as
compared to all previous techniques.

\section{Principal Components Analysis}
\label{sec:pca}

Principal Components Analysis \cite[PCA; see][]{loeve55}, also known as Karhunen-Lo\`eve transformation,
is an algorithm that belongs to the field of
multivariate statistics. Briefly, it is used to obtain a self-consistent basis in
which the data can be efficiently developed. This basis has the property that the 
largest amount of variance is explained with the least number of basis vectors. It is 
useful to reduce the dimensionality of data sets that depend on a huge number of parameters. This
very property has been used during the last decades with denoising purposes, and it
constitutes the main core of the denoising technique that we propose for polarised
spectra.

For the sake of simplicity, we focus from the begining in our particular problem
of polarised spectra. 
When a spectrograph is used to observe a spectral line formed in a stellar
atmosphere, it is sampled at a 
finite number of wavelengths, a number that depends on the spectral resolution
of the instrumental setup. However, it turns out that this number is usually much
larger than the number of physical variables involved in the spectral line
formation mechanism \citep{asensio_dimension07}. Moreover, if we observe the full
Stokes vector, the number of wavelengths increases in a factor of 4, while the number
of physical parameters typically increases more slowly. It is easy
to understand that correlations between the observables exist. This is related to the 
fact that the presence of physical laws constrain the possible values that any
observable can take. For instance, all the wavelength points tracing the continuum away 
from spectral lines roughly provide the same information about the physical conditions.
Since the stellar continuum is typically formed in local thermodynamical equilibrium conditions,
it can be characterized by a Planck function at a given temperature. Therefore, all the
wavelength points are linked by the functional form of the Planck function. 

Due to these intrinsic correlations that exist in the observables, when a spectral line is 
observed many times, or, in our case, several spectral lines are observed simultaneously, the 
cloud of points that represent all spectral lines in the multi-dimensional space of the 
observables will be elongated in some particular directions. These directions are the so-called
principal components and the data can be efficiently reproduced as a linear combination of
vectors along them.

Let us assume that the wavelength variation of each Stokes profile ($I$, $Q$, $U$, or $V$) of a particular 
spectral line is described by the quantity $S_i^j$. The index $i$ represents the wavelength
position while the index $j=\{I,Q,U,V\}$ indicates the Stokes parameter. Each Stokes parameter
is a vector of length $N_\lambda$, corresponding to the number of wavelength points. In the ideal situation, it
would be advantageous to have $N_\mathrm{obs} \gg N_\lambda$ observations, so that the number of 
observed lines is much larger than the 
number of wavelength points used to sample each line. Thanks to the
cross-dispersed capabilities of instruments like SEMPOL, ESPaDOnS or NARVAL, a huge number of
spectral lines is obtained in one exposure when recording spectro-polarimetric data. This fact 
allows us to apply statistical techniques to capture the intrinsic behavior of the 
points in the $N_\lambda$-dimensional space and to use PCA to reduce its dimensionality.

We define $\hat{O}$ as the $N_\mathrm{obs}\times N_\lambda$ matrix containing the wavelength variation 
of all the observed spectral lines. The principal components can be found as the 
eigenvectors of this matrix of observations.
This means that the PCA procedure reduces to the diagonalisation of the matrix
$\hat{O}$. Since we are requiring that $N_\mathrm{obs} \gg N_\lambda$ holds, this matrix is not square by 
definition. Moreover, even if one uses the Singular Value Decomposition 
\cite[SVD; see, e.g.,][]{numerical_recipes86} to diagonalise $\hat{O}$,
the dimension of the matrix can be so large that
computational problems can arise. It can be demonstrated that the right singular vectors of 
of the matrix $\hat{O}$ are equal to the singular vectors of the cross-product matrix:
\begin{equation}
\hat{X}=\hat{O}^t\hat{O}.
\label{ec1}
\end{equation}
The matrix $\hat{X}$ is the $N_\lambda \times N_\lambda$ cross-product matrix and the superindex
``t'' represents the transposition operator. The same applies to the left singular
vectors, which are also eigenvectors of the cross-product matrix $\hat{X}'=\hat{O}\hat{O}^t$.
The matrix $\hat{X}'$ has dimensions $N_\mathrm{obs} \times N_\mathrm{obs}$ and is typically much larger than 
the matrix $\hat{X}$. However, one description is the dual of the other and they are 
completely equivalent. The $i$-th principal component, $\vec{B}_i$, fulfills:
\begin{equation}
\hat{X} \vec{B}_i = k_i \vec{B}_i,
\end{equation}
with $k_i$ its associated eigenvalue. All the eigenvectors can be put together in the
matrix $\hat{B}$. This matrix has dimensions $N_\lambda \times N_\lambda$
and contains the eigenvectors as column vectors. Note that the cumulative distribution of
eigenvalues
\begin{equation}
g_m = \frac{\sum_{i=1}^m k_i}{\sum_{i=1}^{N_\lambda} k_i}
\label{eq:cumulative_var}
\end{equation}
gives the relative amount of variance explained by the first $m$ eigenvectors.
Since these vectors constitute a basis, the observations can be
written as a linear combination of them as follows:
\begin{equation}
\hat{O}=\hat{C}\hat{B}^{t},
\label{ec2}
\end{equation}
being $\hat{C}$ the $N_\mathrm{obs}\times N_\lambda$ matrix of coefficients. The element $C_{ij}$
of this matrix represents the projection of the observation $i$ on the eigenvector $j$. This
matrix can be easily calculated as:
\begin{equation}
\hat{C}=\hat{O}\hat{B}.
\label{ec3}
\end{equation}
Note that the transposition operator of the matrix of the eigenvectors in Eq. (\ref{ec2})
replaces the inverse operator because the matrix of singular vectors is orthogonal, so that
it fulfills $\hat{B}^{-1}=\hat{B}^{t}$. This greatly simplifies the calculations because no
numerical matrix inversion is needed.

\begin{figure*}[!t]
\includegraphics[width=0.33\textwidth]{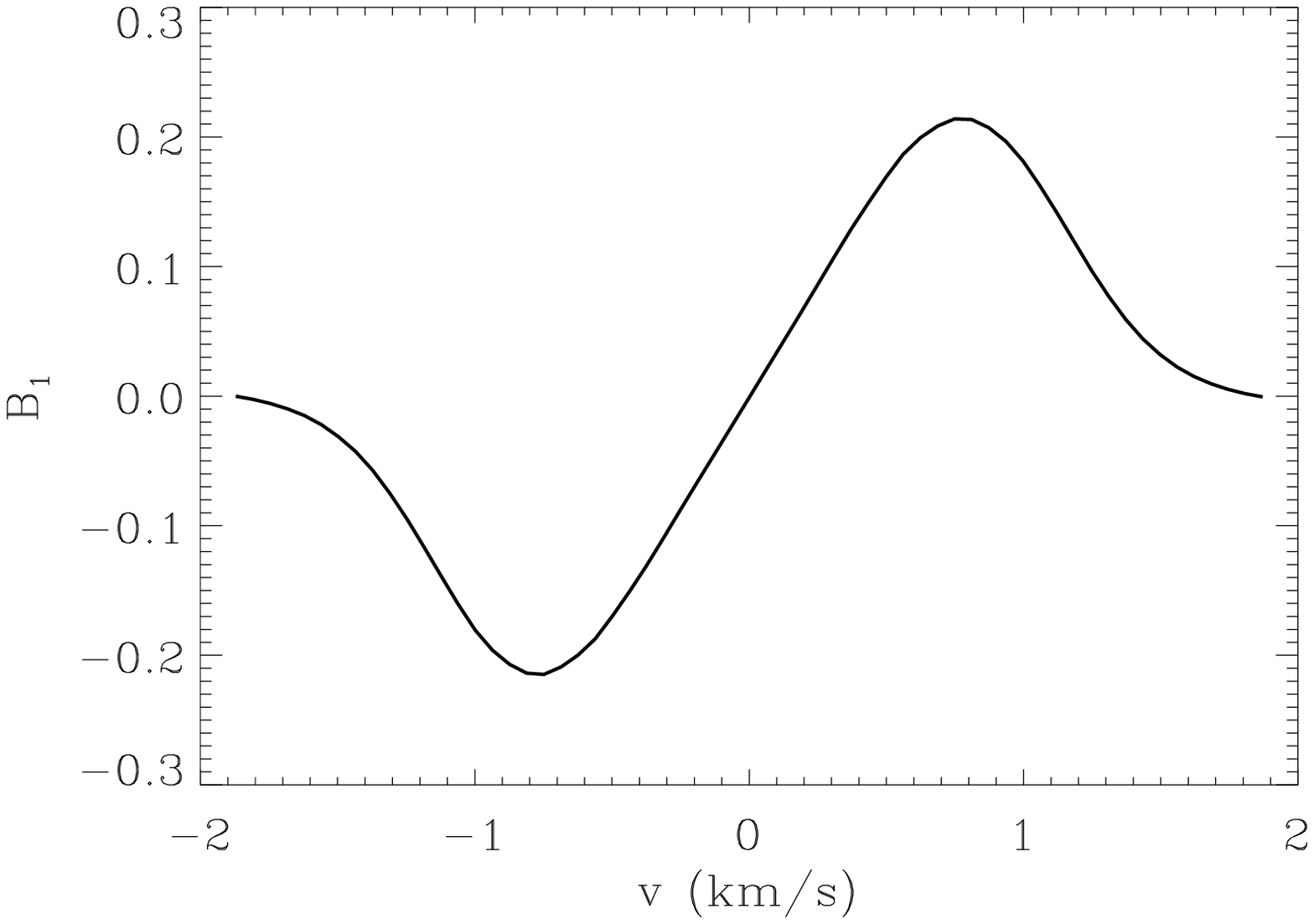}
\includegraphics[width=0.33\textwidth]{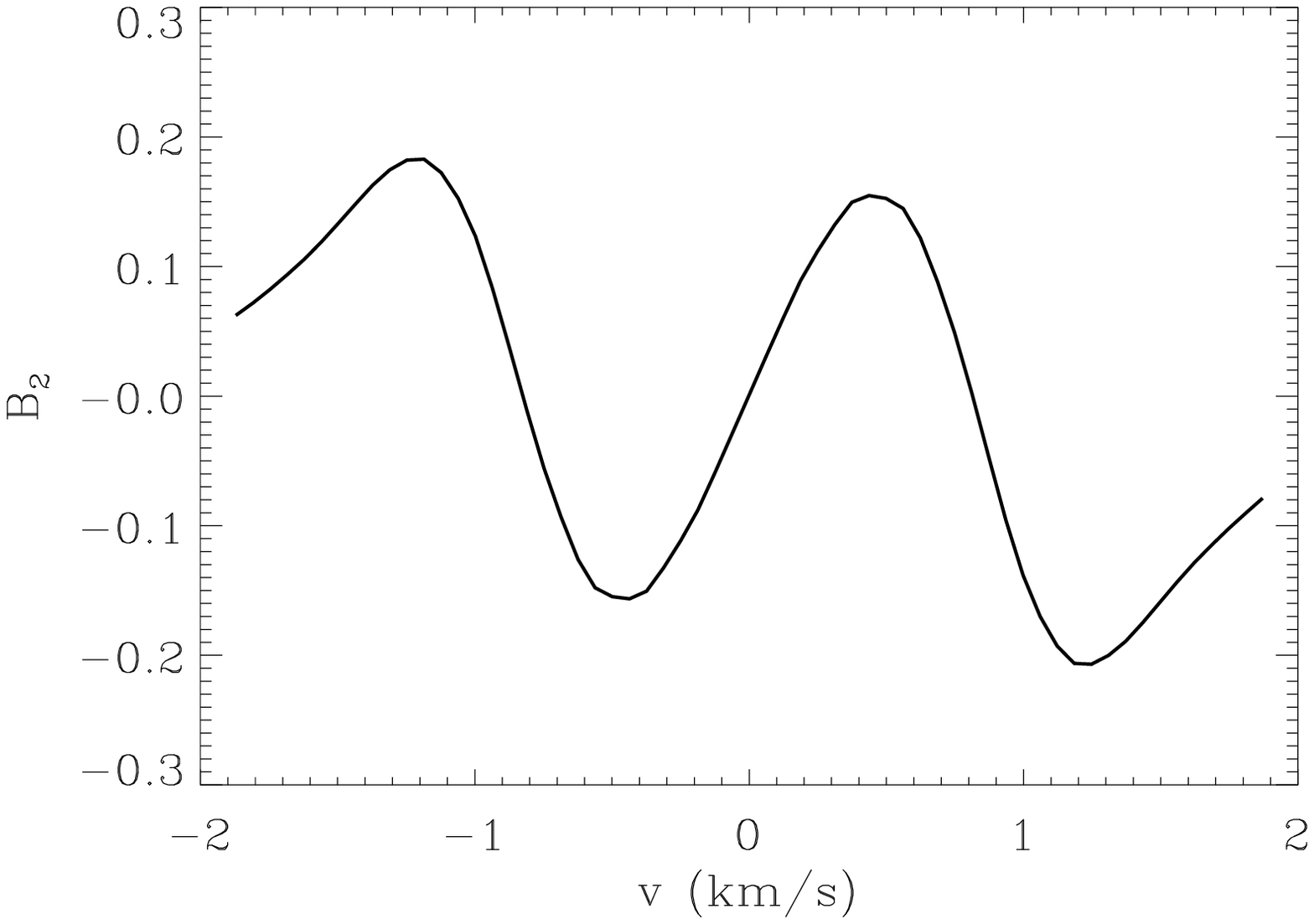}
\includegraphics[width=0.33\textwidth]{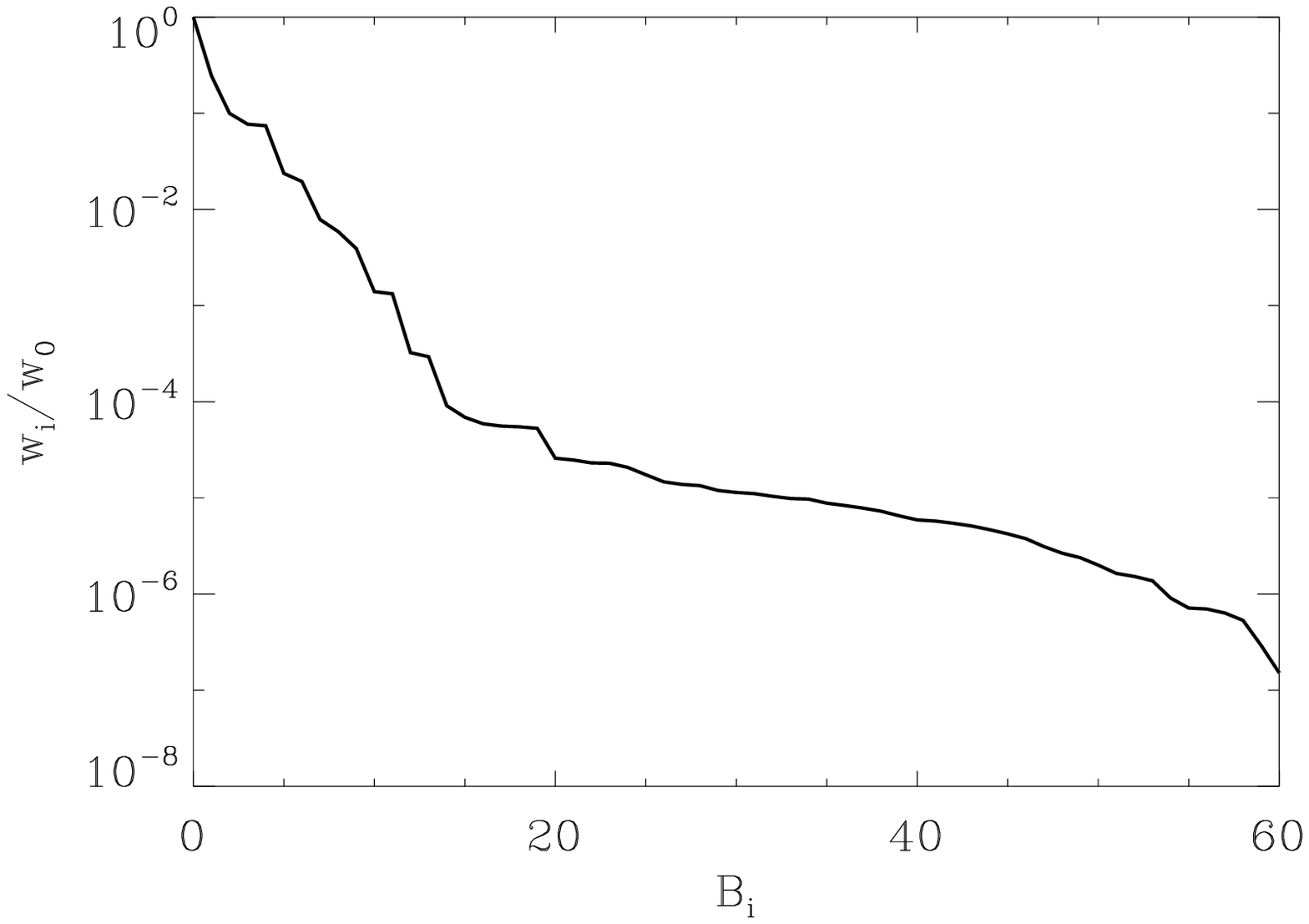}\\
\includegraphics[width=0.33\textwidth]{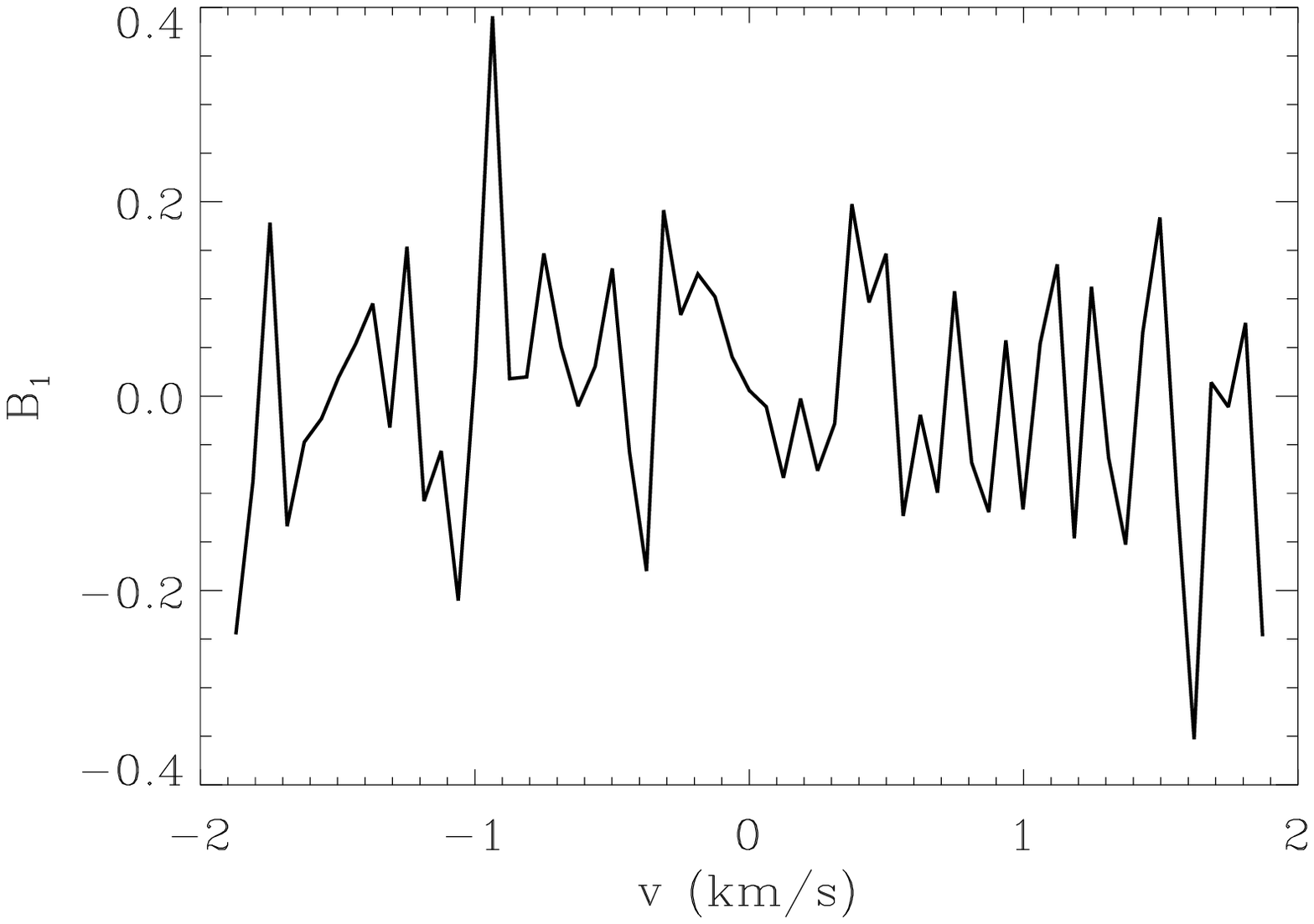}
\includegraphics[width=0.33\textwidth]{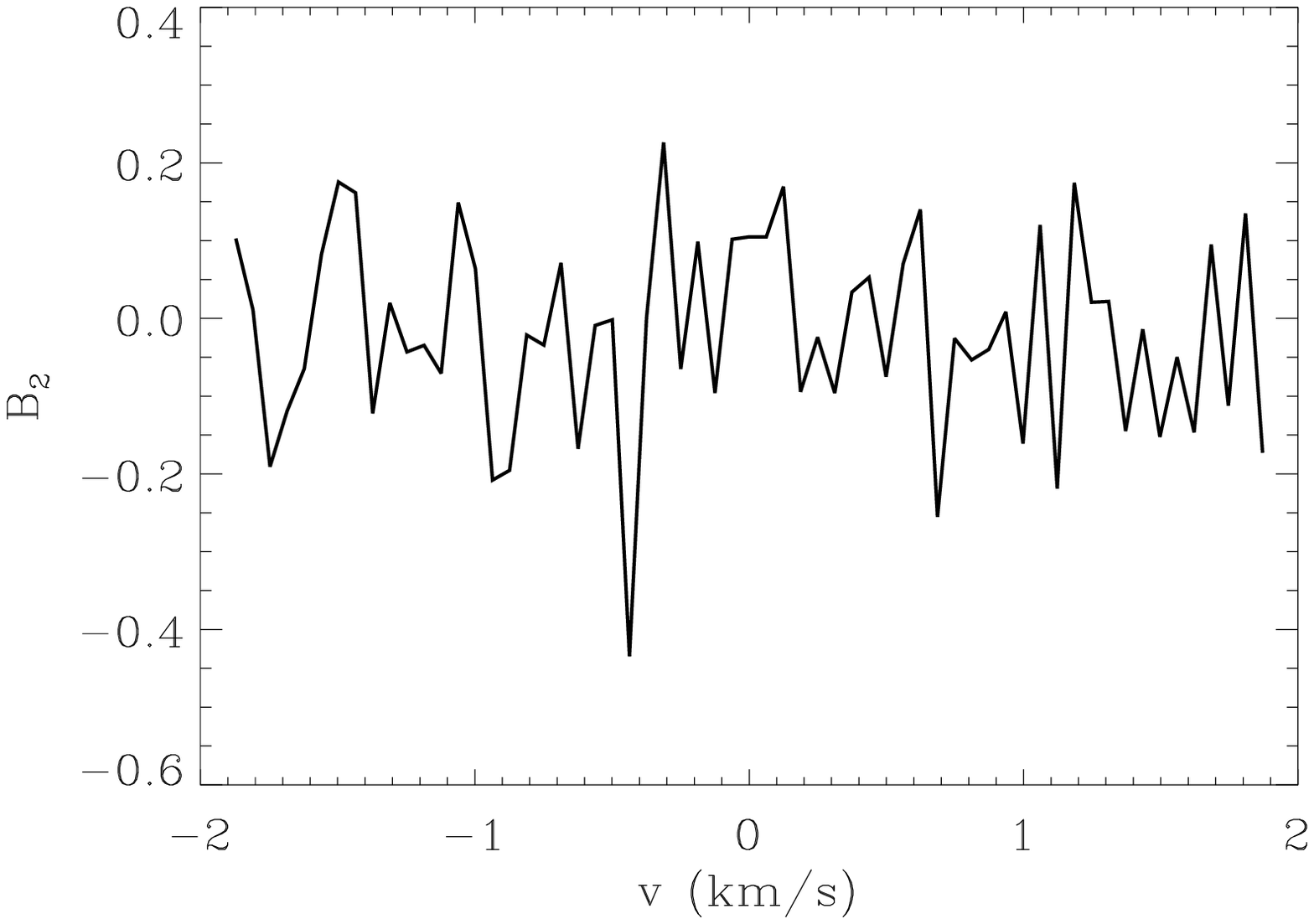}
\includegraphics[width=0.33\textwidth]{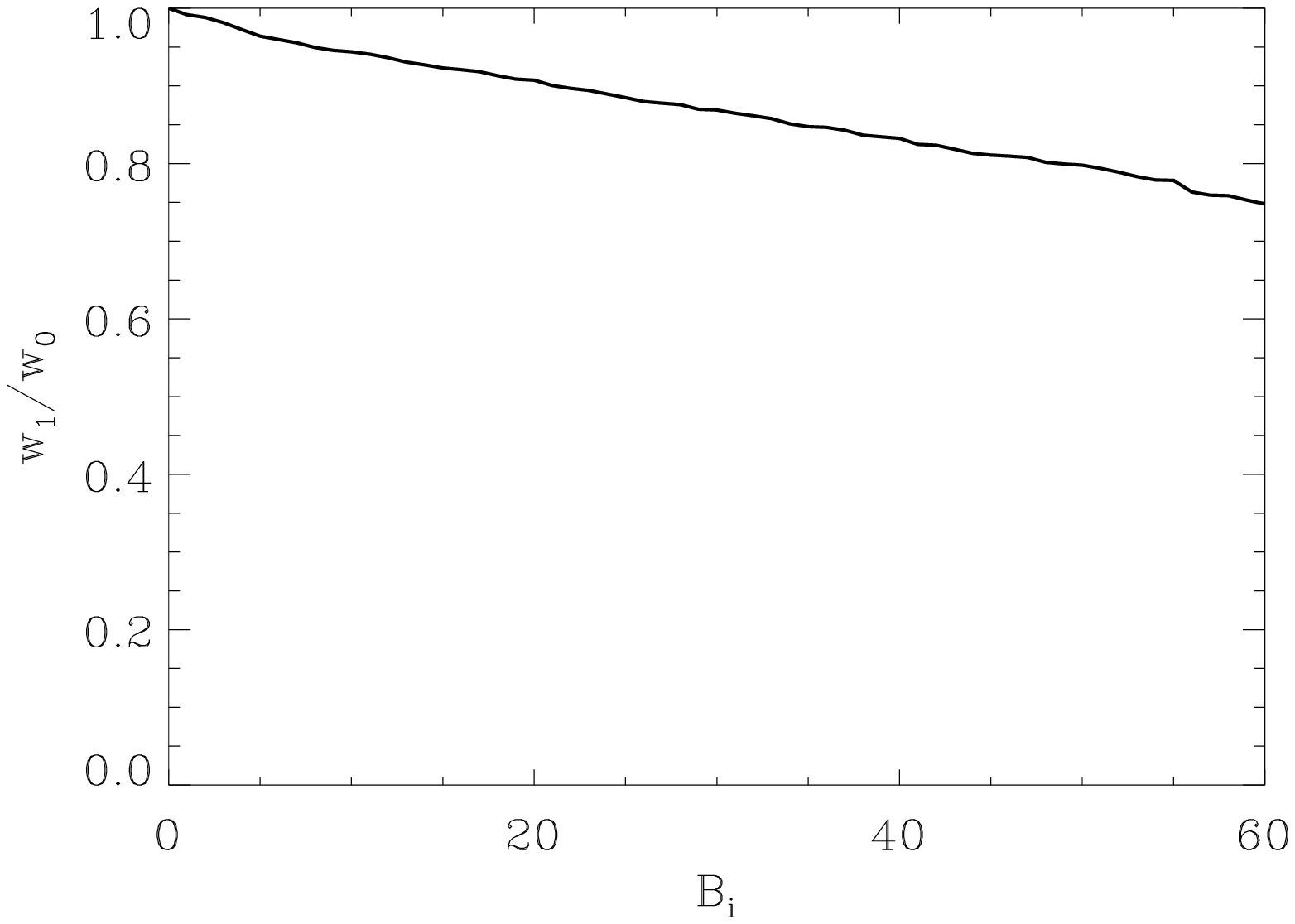}
\caption{Top panels: first (left) and second (centre) eigenvectors of the matrix
of observations without any noise added. The dataset is described in Section \ref{sec:dataset}.
The right upper panel represents the eigenvalues
for all the eigenvectors. Bottom panels: equivalent to the upper panels but for
an observation containing only gaussian noise. The standard deviation of the distribution of noise is $10^{-3}$
I$_\mathrm{c}$.}
\label{autovect}
\end{figure*}

\begin{figure*}[!ht]
\includegraphics[width=\columnwidth]{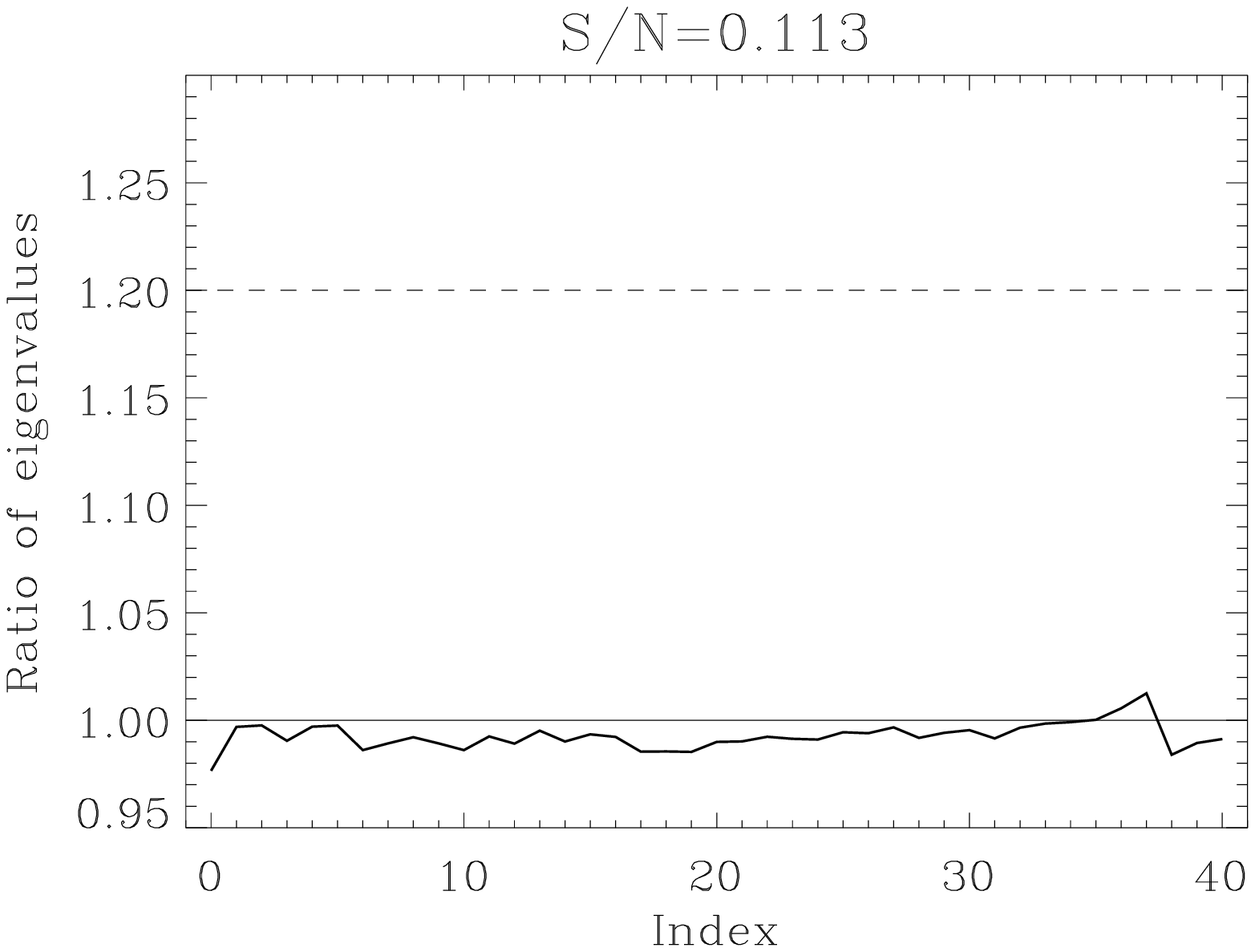}
\includegraphics[width=\columnwidth]{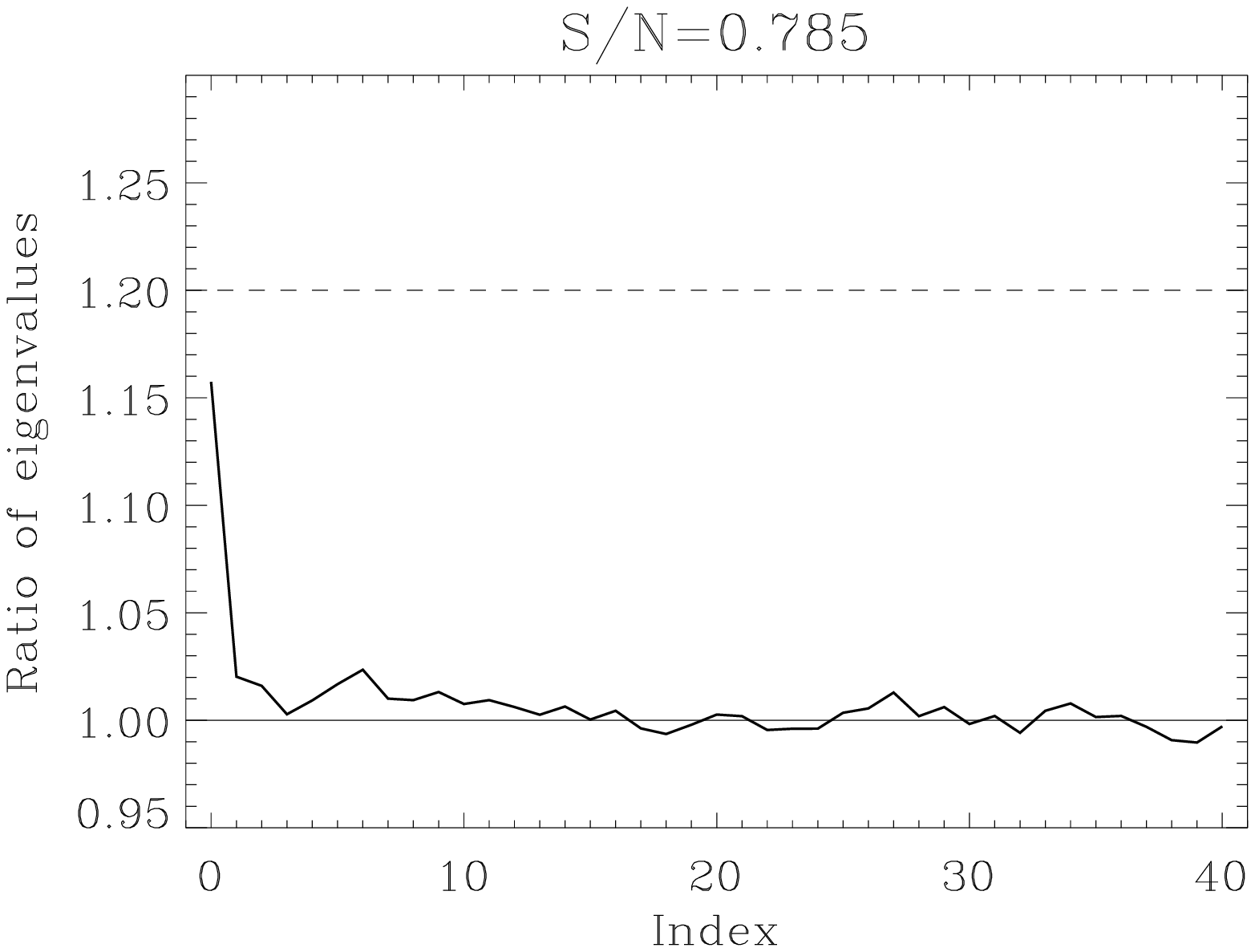}
\includegraphics[width=\columnwidth]{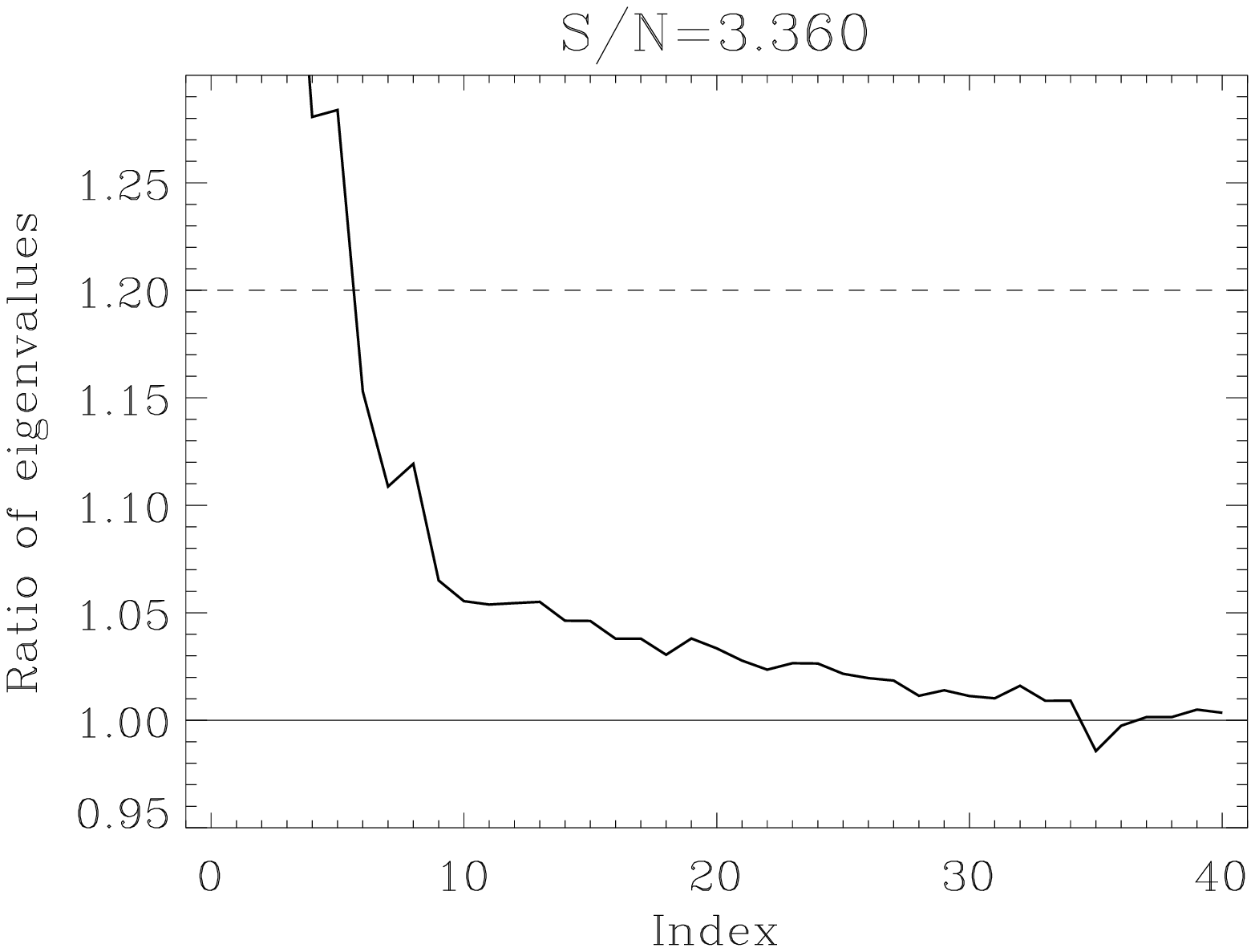}
\hspace{0.35cm}\includegraphics[width=\columnwidth]{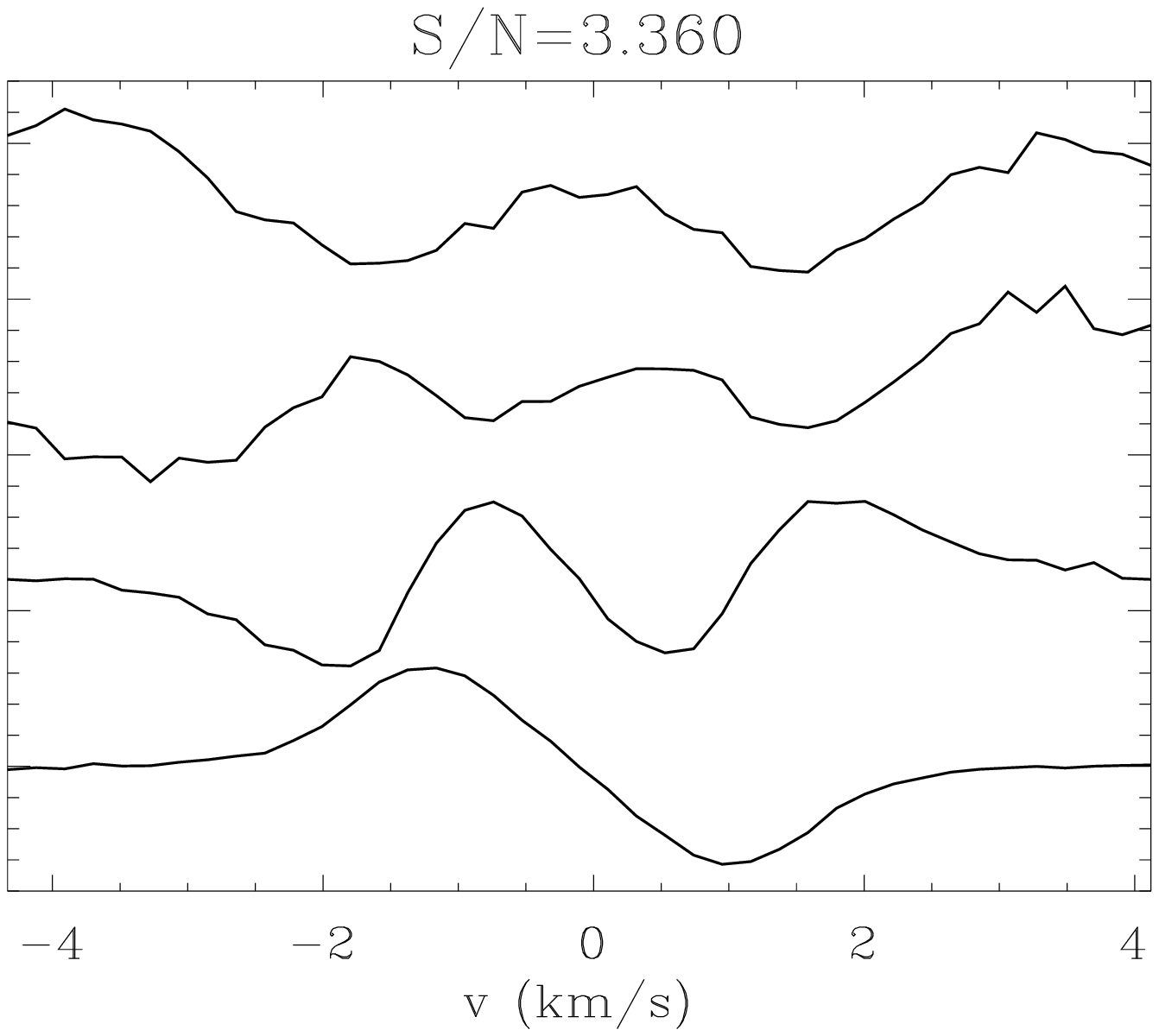}
\caption{Ratio between the eigenvalues of the noisy observations obtained for different S/N values and the ones of the pure noise observations (top and bottom left panels). The horizontal dashed line indicates the value $f=1.2$.
The bottom right panel shows the first four eigenvectors (ordered from bottom to top) for the case 
of $S/N=3.360$. The eigenvectors are shifted in the vertical direction to avoid overlapping.}
\label{autoval_ratio}
\end{figure*}

\section{Denoising procedure}
In the following, we explain the procedure that we propose for denoising the
experimental Stokes profiles of magnetic stars. The method is general, it should be used not 
only to retrieve Stokes I and V but to
obtain Stokes $Q$ and $U$ given the important information encoded in it 
\citep[see, e.g., the series of paper starting with][]{landolfiI93}.

\subsection{The simulated data set}
\label{sec:dataset}
In order to demonstrate the capabilities of the PCA denoising technique, we use
a synthetic polarised spectrum, including Stokes $Q$, $U$ and $V$. We cover the wavelength range between
400 and 900 nm, with a spectral resolution of 50 m\AA. The synthetic spectrum has been 
obtained under the assumption of local thermodynamical equilibrium (LTE) using a standard
solar model atmosphere \citep{fontenla_falc93}, with a star-filling magnetic field of 1000 G. 
The inclination of the magnetic field with respect to the line of sight is 45$^\circ$ and its inclination is 
20$^\circ$. This produces polarization signals that are much larger than those observed in real
cases. For this reason, we apply a filling factor $f$ to our simulated spectra in order to end up
with Stokes $V$ amplitudes that are similar to the ones expected in some cool stars 
observations ($\sim 10^{-4} I_\mathrm{c}$, being $I_\mathrm{c}$ the continuum intensity). 
We quantify the quality of the data (amount of information about the physical conditions
in the regions of line formation available in the data) with the signal to noise ratio, $S/N$. 
Consequently, the filling factor turns out to be
unimportant and it is only chosen so as to end up with amplitudes comparable to the observed ones.
The influence of realistic surface magnetic field distributions on the capabilities of the
denoising technique will be addressed by Carroll et al. (2008; in preparation).

The spectral range that we use in our denoising technique is very large (500 nm), so that
there is a large difference between the Doppler width of lines in the red and in the blue
part of the spectrum. This is because the Doppler width is proportional to the wavelength.
In order to make the Doppler widths compatible for all wavelengths, we transform
the wavelength axis into the following velocity axis:
\begin{equation}
v=c\log{\frac{\lambda}{\lambda_\mathrm{ref}}},
\end{equation}
where $c$ is the speed of light, the symbol $\lambda$ represents
the wavelength and $\lambda_\mathrm{ref}$ is a reference wavelength, which we choose to
be 400 nm. This change of variables induces that all the lines have, to first order, the same
Doppler width in the new axis. Differences may exist because the Doppler width depends on the 
atomic mass of each species and because it also depends on the temperature in the line formation
region. However, we assume that these differences are of second order with respect to the 
wavelength dependence. Since this new axis has an irregular step size because the spectrum has been
sampled regularly in a wavelength scale, we re-interpolate it to a velocity axis with a regular 
step using a standard linear interpolation procedure. We set the spectral resolution equal to 0.2 km/s.
This is equivalent to assume that the spectral resolution is the same regardless of the wavelength.

The individual spectral lines that will be used for building the matrix $\hat{O}$ will
be extracted using fixed positions for the central wavelength. In this experiment, 
we have computed the positions of the spectral lines as the positions where the minimum
of the intensity profile is found. In any case, 
standardized linelists have been developed for different stars depending on the spectral type
\citep{donati97}. The results that we show in this paper have been obtained using
a database with $\sim$6300 lines. In principle, the capabilities of the method might
be improved by using databases with more spectral lines, provided that the added lines
carry sufficient information.
We set $N_\lambda=40$, choosing 20 points to the red and 20 points to the blue.
This translates into a velocity range of \hbox{8 km/s}, which is sufficient for our experiment
since we are not including the effects of rotation. In any case, in the analysis of a rapid
rotator, a larger number of points have to be chosen. With each individual profile, we construct the
matrix of observations ($\hat{O}$) having one spectral line in each one of the
rows.

\subsection{Principal components of a ``correlated'' data set}
We refer to a ``correlated'' data set when some correlation between the
observables exist. In our particular case, this means that the physical mechanisms
of line formation in stellar atmospheres introduce correlations between 
different wavelenght points of each spectral line. The principal components of a
correlated data set have some peculiarities that allow us to reduce the
dimensionality of the data set. The principal components associated with the largest
eigenvalues are representative of the directions of highest correlation and 
Eq. (\ref{eq:cumulative_var}) can be used to estimate the relative amount of variance
explained by them. Top panels of Fig. \ref{autovect} show the
first two eigenvectors of the matrix of observations of the Stokes V parameter without any noise added.
The first eigenvector has the typical antisymmetric shape representative of the Stokes $V$ profile
induced by the Zeeman effect. This means that the most important common pattern to all of our
spectral lines ressembles a Zeeman profile. Note also that the first eigenvalue is much larger
than the following ones (right panel of Fig. \ref{autovect}; note the logarithmic scale). Although this is an expected result, note 
that we have not assumed in the analysis any systematic pattern in our data. On the 
contrary, it is a natural result of PCA. The rest of eigenvectors present
other characteristics of the profiles whose importance decreases as the associated eigenvalue 
decreases.

The right panel of Fig. \ref{autovect} shows that the first eigenvalues is the most 
representative one and that they drop dramatically. This is the key property of the PCA
that allows us to reduce the dimensionality of the data set. This means that
our observations can be efficiently reproduced using only a few eigenvectors. The observations
were represented in a space of $N_\lambda$ dimensions but the PCA analysis indicates that
it is possible to represent them in a space of $N_\lambda'$ dimensions (the number of
chosen eigenvectors), with $N_\lambda' \ll N_\lambda$.

\subsection{Principal components of uncorrelated noise}
It is instructive to apply the same analysis based on principal components to a data set 
composed only of uncorrelated noise. In the limit $N_\lambda \to \infty$ and 
$N_\mathrm{obs} \to \infty$, the cross-product matrix is strictly equal to the identity. 
As a consequence, the eigenvectors are the canonical basis and the eigenvalues are all 
equal to 1. However, since $N_\lambda$ is small, the cross-product matrix has non-diagonal
elements and some spurious correlation may appear between different wavelength points. 
The bottom left and middle panels of Fig. \ref{autovect} show the first
two eigenvectors of a matrix with the same size as the one of the
observations but containing only gaussian noise\footnote{Note also that the random numbers obtained
in computers are not strictly uncorrelated and this can induce (hopefully small) additional non-zero non-diagonal
elements in the cross-product matrix.}. The gaussian distribution of noise has a
standard deviation equal to $10^{-3}$ I$_\mathrm{c}$. All the eigenvectors are
noisy, similar to those shown in the figure. The eigenvalues, represented in the 
bottom right panel of Fig. \ref{autovect} are roughly the same for all the eigenvectors.

Real spectro-polarimetric measurements can present some degree of correlation between
different wavelength points but are also contaminated by uncorrelated noise. Consequently, it
is clear that the noise level will be an important issue that will restrain the
denoising of the observations by means of principal components.

\begin{figure}[!t]
\includegraphics[width=\columnwidth]{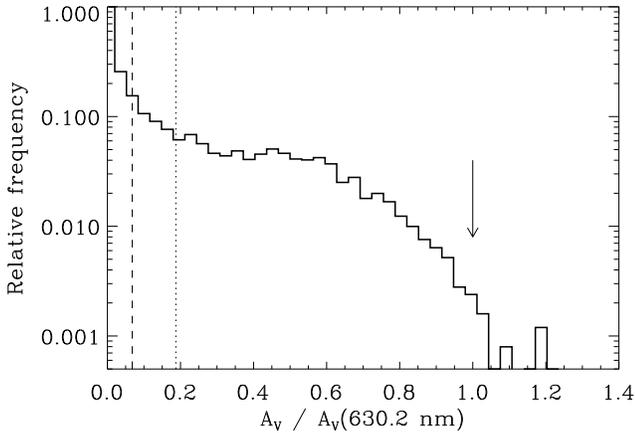}
\caption{Histogram of the amplitudes of the Stokes V profiles in the blue lobe divided the amplitude of the 630.2 nm spectral line. 
The histogram has been normalized to its maximum. The dashed line marks the position of the median value of the amplitudes while the 
dotted line shows the mean value. The arrows points to the value of 1 that corresponds to the 630.2 nm spectral line.}
\label{ampl_v}
\end{figure}

\subsection{The procedure}
\label{sec:procedure}
To simulate the real case, we add gaussian noise to each spectral line of our
matrix of observations. We apply the denoising procedure to our simulated
data set with different values of added noise in order to analyze how it behaves
in different situations. In order to use only one reference signal to noise ratio 
we choose the ratio between the polarisation signal ($Q$, $U$ or $V$) in the magnetically sensitive 630.2 nm line 
and the standard deviation of the added noise distribution. Note that this $S/N$ gives an idea of
the quality of the data for the most magnetically sensitive lines, while the median of the
$S/N$ distribution is located at a much lower value, typically one order of magnitude
smaller. This is produced by the fact that most of the lines induce small polarization
signals, while only few lines give conspicuous signals. Consequently, the distribution of
 amplitudes is strongly shifted towards zero.

For a given $S/N$, the eigenvectors of the cross-product matrix are computed as described in
Section \ref{sec:pca}. If any systematic Zeeman signature is present in the dataset, it 
will appear in the first few eigenvectors. Since uncorrelated noise is also present in the
observations (perhaps even completely masking the line polarization signals), the rest of eigenvectors 
having smaller eigenvalues contain the contribution of this noise. The filtering procedure 
consists on reconstructing the observed signal using only the first eigenvectors:
\begin{equation}
\hat{O'}=\hat{C'}\hat{B'}^{t},
\end{equation}
where $\hat{O'}$ is the matrix of observations after the denoising procedure.
The matrix $\hat{B'}$ contains the few first eigenvectors that have been retained as 
containing stellar magnetic signatures. The matrix $\hat{C'}$ contains the coefficients of
the projection of the matrix of observations onto the chosen basis of eigenvectors:
\begin{equation}
\hat{C'}=\hat{O}\hat{B'}.
\end{equation}

The selection of the number of eigenvectors that are dominated by polarimetric
signal is the fundamental free parameter of the denoising procedure. It is not an easy task
to efficiently select it and sometimes requires subjective criteria. We have verified
that the following criterium works quite well for many of the tested cases and
is also based on the properties of the PCA decomposition. For each value of the
variance of the noise added to the simulated profiles, we compute an observation matrix
equal to the observed one but made only of uncorrelated noise. In real situations, this
matrix has to be built based on the estimation of the noise present in the 
observations. The eigenvalues of the pure noise matrix are calculated and 
compared with those of the real observations. If the variance of the noise has been
correctly estimated, the two eigenvalue distributions will overlap except for those
eigenvalues associated with correlated signals. Consequently, we select those eigenvectors
$b_i$ with eigenvalues higher than a factor $f$ of those corresponding to the pure
noise case. We have verified that $f \approx 1.2$ gives good results. It is also instructive
not to rely on automatic selection methods but verify the shape and weight of each
eigenvector. The direct analysis of the eigenvectors can show many important details about
the hidden signal and some tricks can be used to enhance the possibility of recovering it.
A detailed discussion of the properties of the eigenvalues and the choice of the cutoff for the 
SVD problem can be found in \cite{christensen93} and in a series of papers by \cite{hansen92}, \cite{hansen_sekii92} and \cite{hansen93}.

It is important to point out that the results we are presenting here are not the 
optimal case. For a field of 1000 G and since we do not include any additional line
broadening mechanism, the majority of the lines are not in the weak field regime of
the Zeeman effect. Therefore, they present different shapes (as a consequence of
the Zeeman saturation) and part of the correlation is lost. We show that the
PCA denoising technique is very performant in this non-optimal case. In stars with
broadened lines (due to whatever mechanism), the weak field regime can be expected
for larger magnetic field strengths and our PCA denoising algorithm will work even
better.

In case none of the
eigenvalues fulfill the criterium (to be expected in extremely noisy spectra), we have chosen
to reconstruct using only the first eigenvector. The upper and bottom left panels of Fig. \ref{autoval_ratio} show, for the cases 
discussed in the next section, the ratio
between the eigenvalues of the cross-product matrix obtained from the synthetic data plus noise
and the eigenvalues of the pure noise cross-product matrix. The horizontal dashed line indicates
the threshold that we choose to select the eigenvectors. The bottom right panel of the figure 
shows the first four selected eigenvectors for the less noisy case that we analyze in this paper.

\section{Results}
We present in this section the behavior of the PCA denoising in several $S/N$ regimes. We
range from very noisy profiles in which the signal is completely masked by the noise
to less noisy profiles in which the PCA technique can be used to improve even more the
quality of the data for the analysis of individual spectral lines. For the sake of 
simplicity, all the figures showing individual line profiles present results for Stokes 
$V$, although similar results (for
similar values of the $S/N$) are obtained for Stokes $Q$ and $U$. However, the general denoising
trends are presented both for circular and linear polarisation states.

\begin{figure}[!t]
\includegraphics[width=0.49\columnwidth]{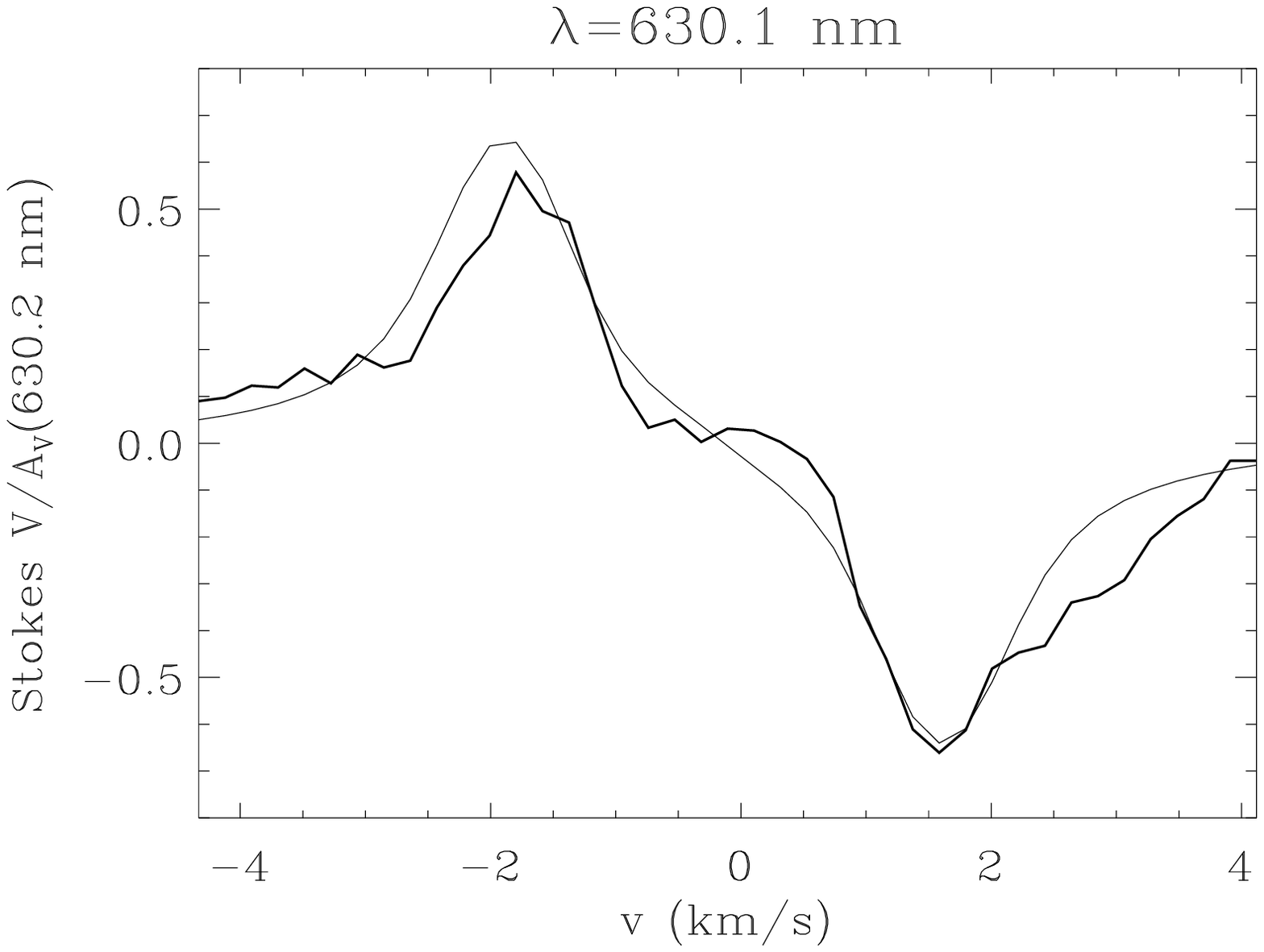}
\includegraphics[width=0.49\columnwidth]{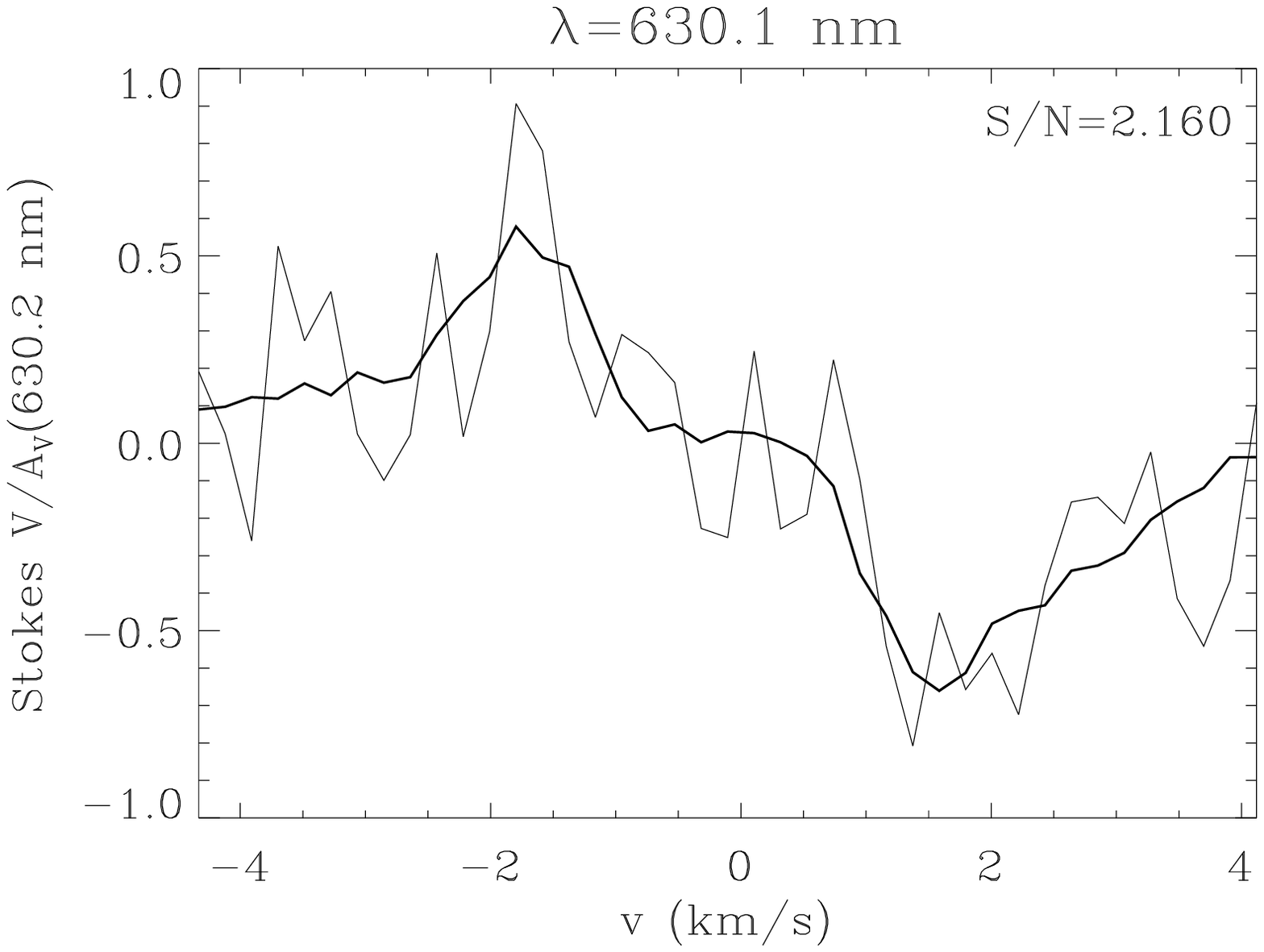}\\
\includegraphics[width=0.49\columnwidth]{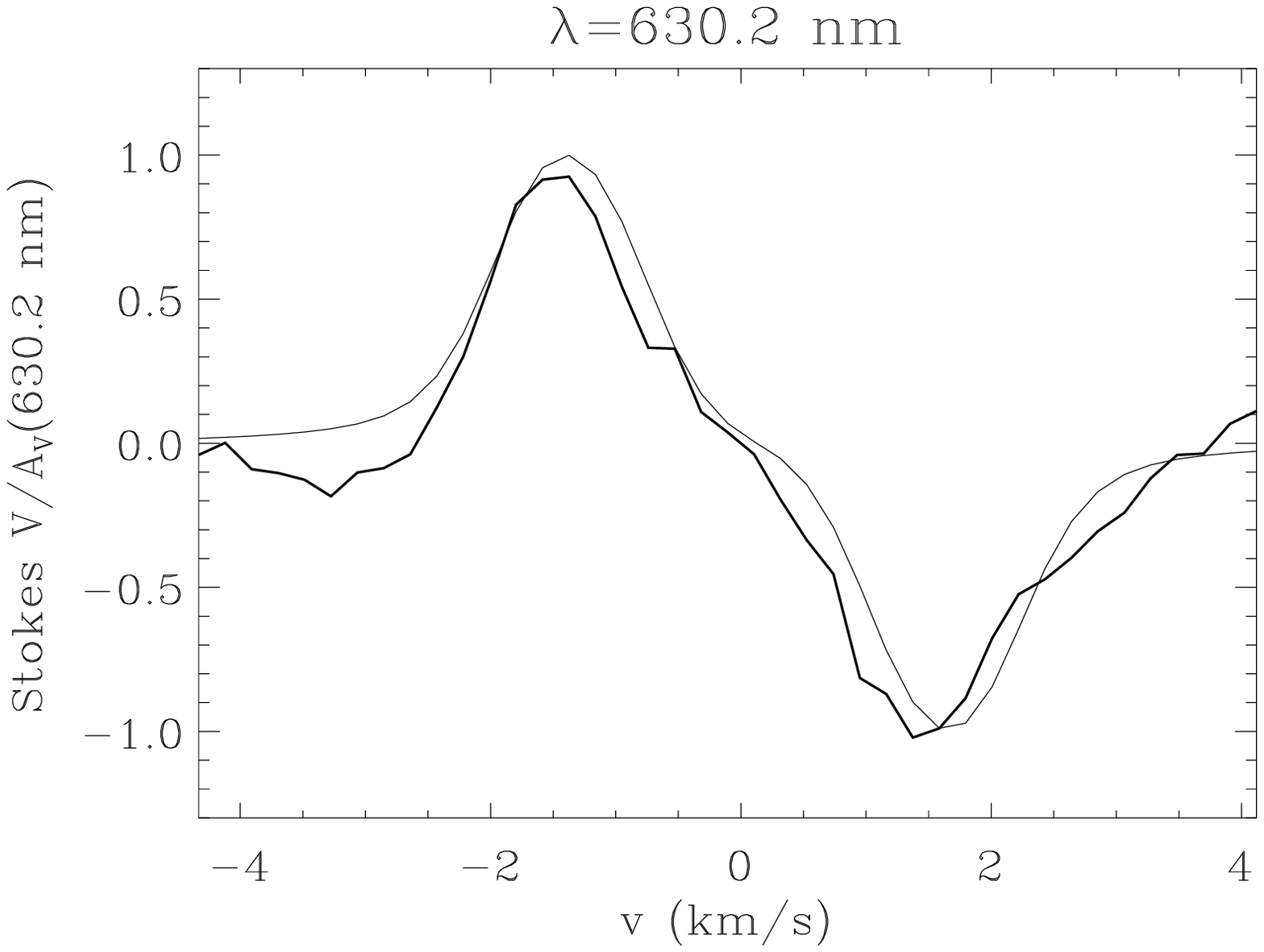}
\includegraphics[width=0.49\columnwidth]{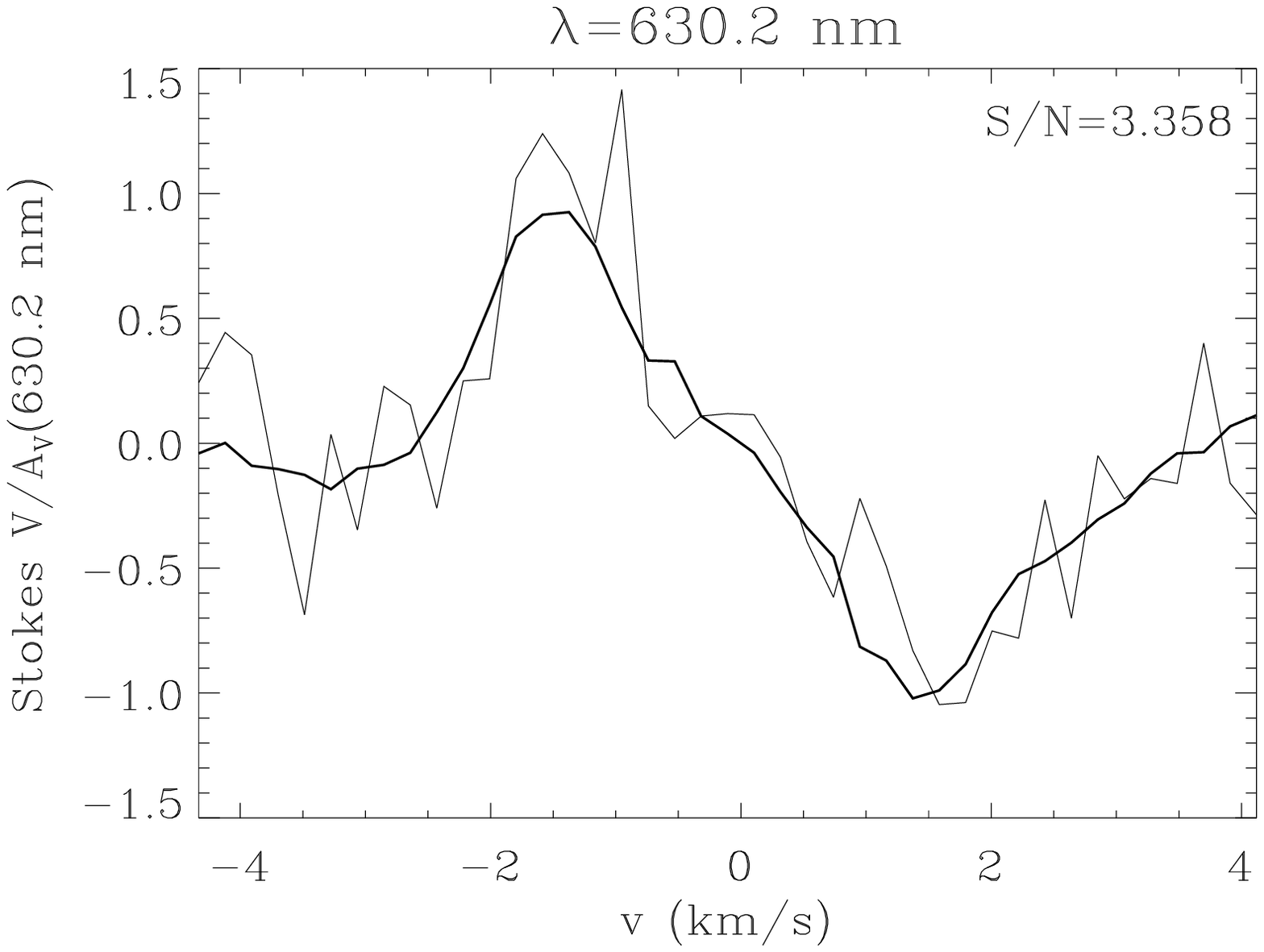}\\
\includegraphics[width=0.49\columnwidth]{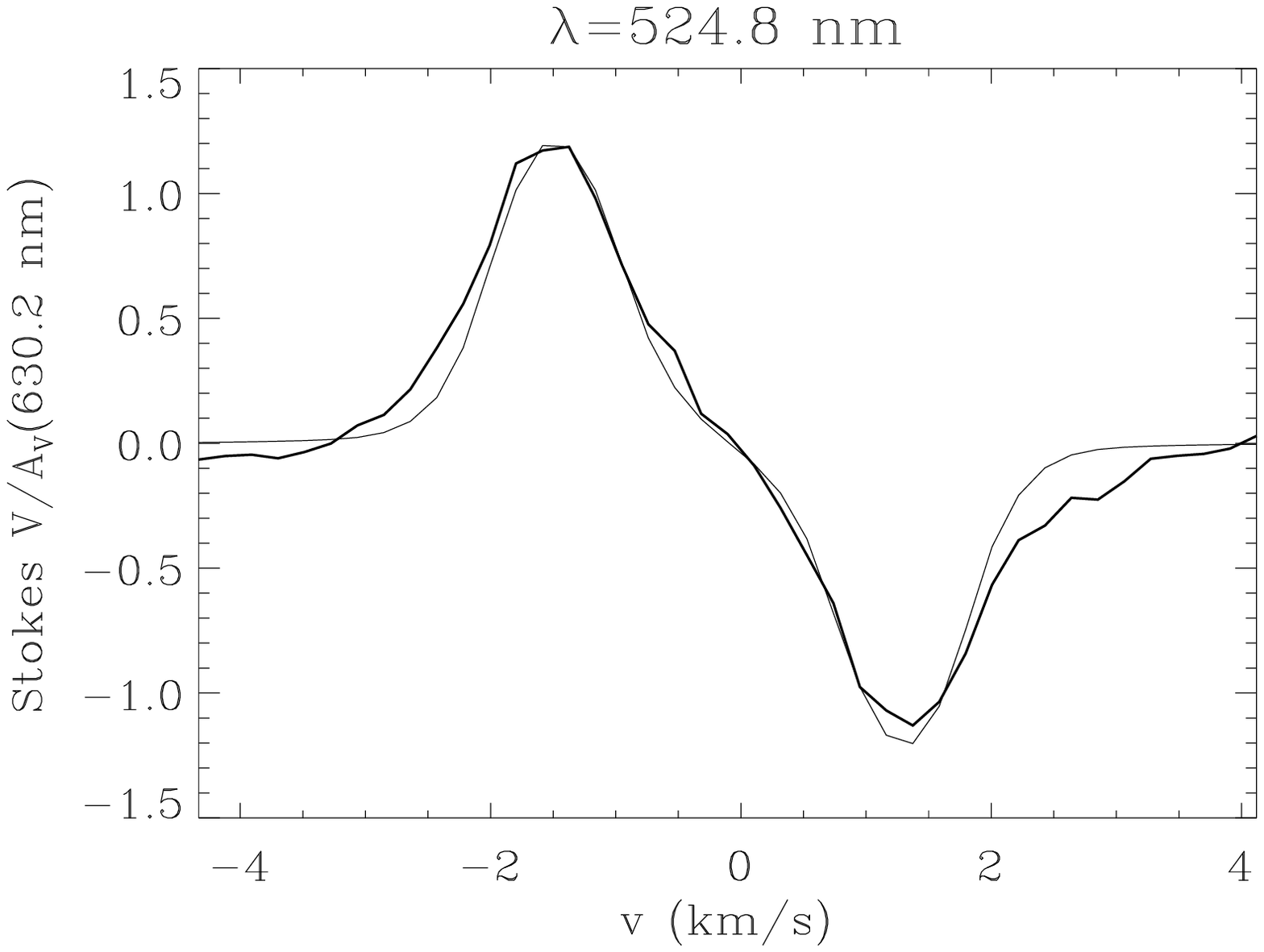}
\includegraphics[width=0.49\columnwidth]{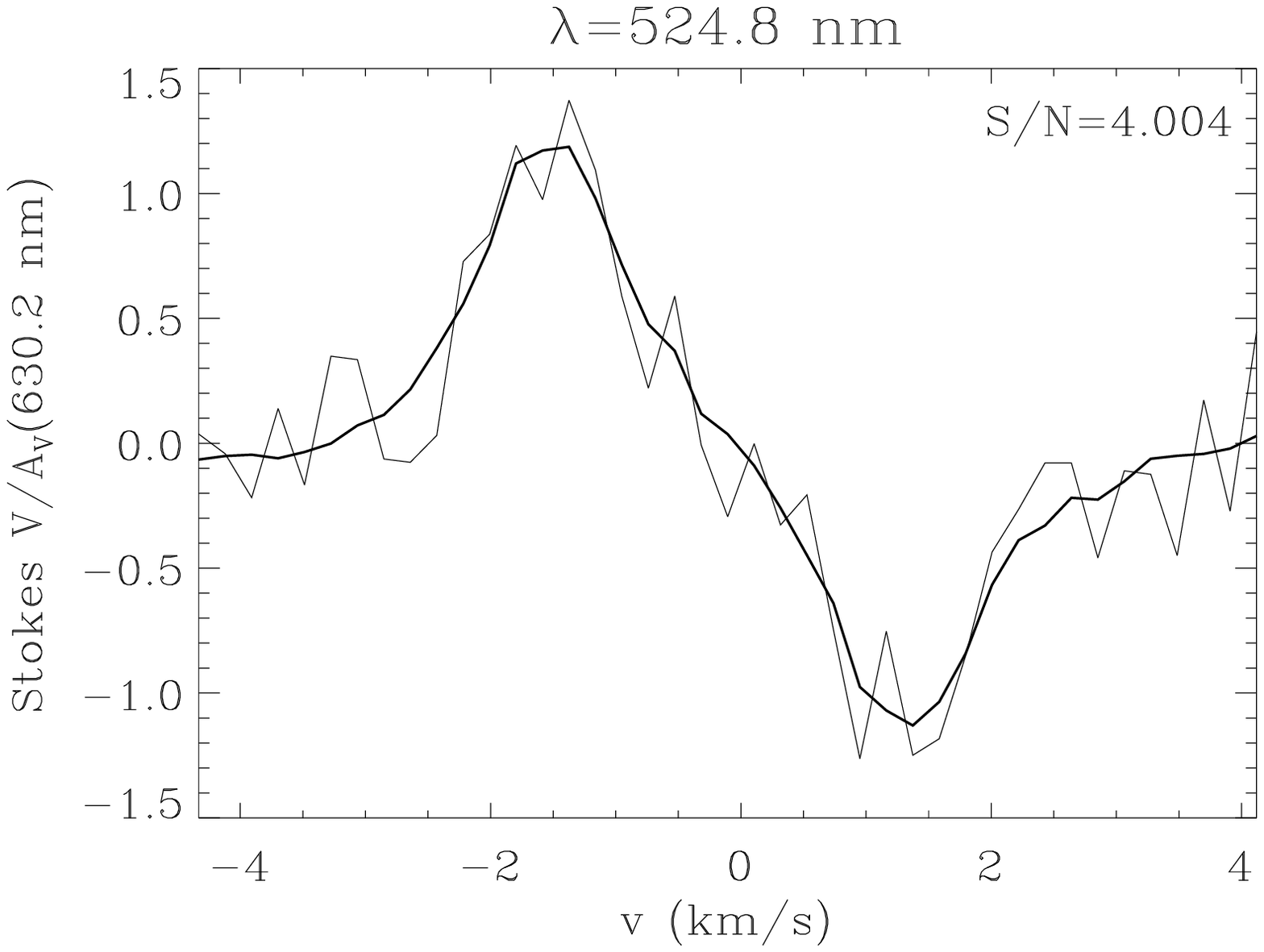}\\
\includegraphics[width=0.49\columnwidth]{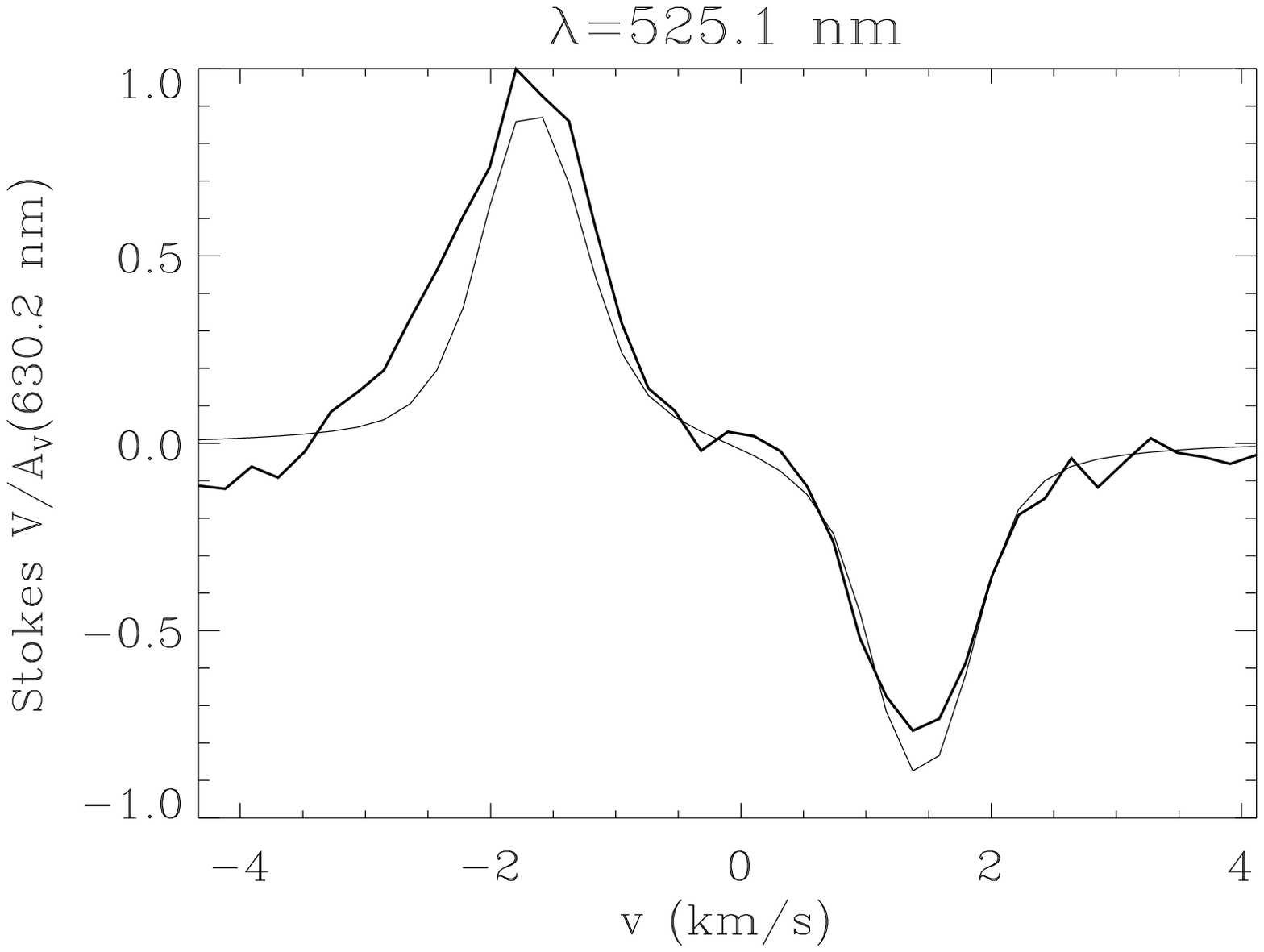}
\includegraphics[width=0.49\columnwidth]{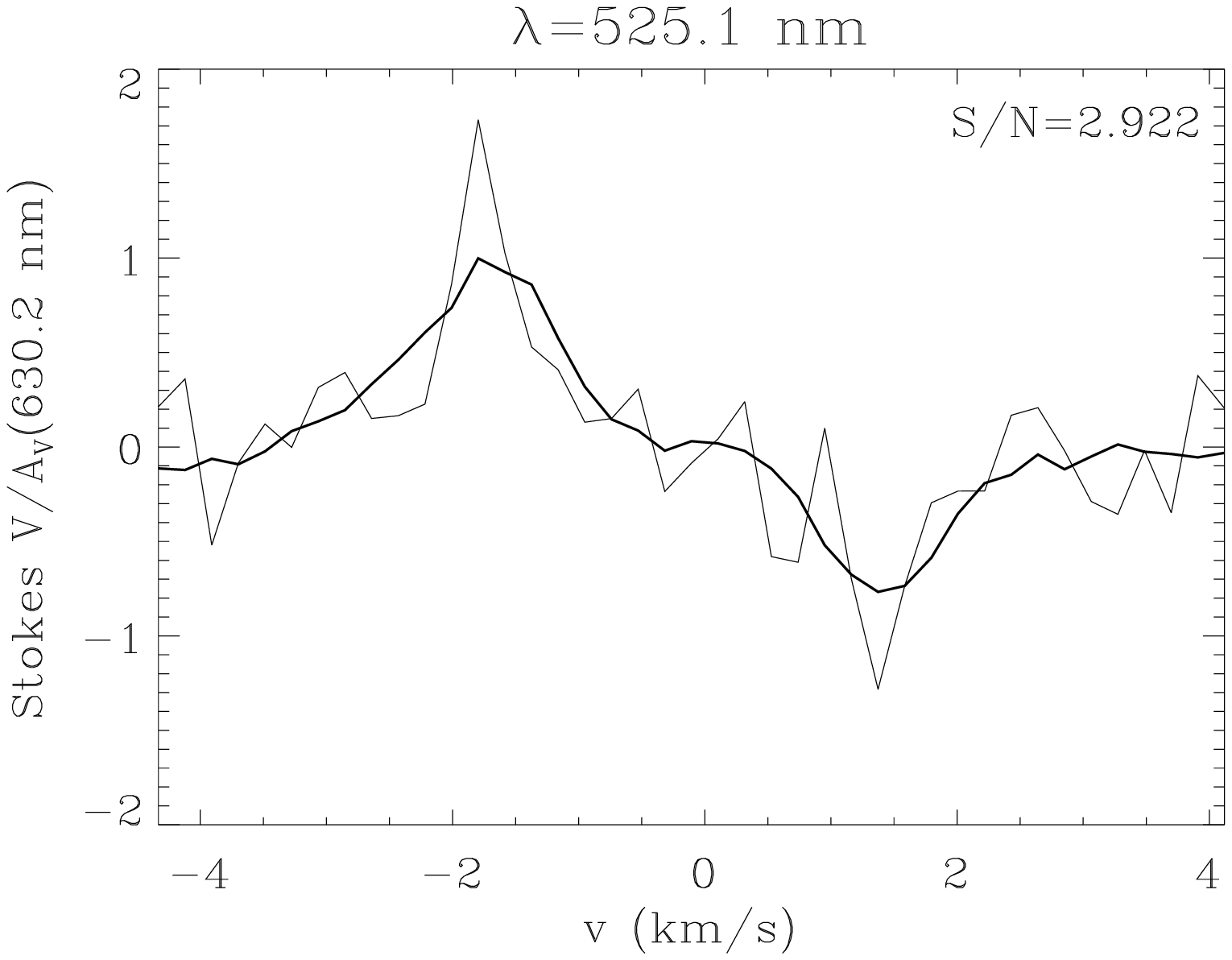}
\caption{Each row of this figure correspond to a particular spectral line. The
first one is the Fe\,{\sc i} line at 630.1 nm, the second one is the 
Fe\,{\sc i} spectral line at 630.2 nm, the third one is the
Cr\,{\sc i} line at 524.8 nm, and the fourth one is the Fe\,{\sc i} line at 525.1 nm. 
The amplitudes are normalized to the amplitud at the blue lobe of the 630.2 nm spectral line. 
Left panels represent the
original profiles of each spectral line (thin line) and the filtered profiles
(thick line). Right panels represent the noisy profile of each spectral line
(thin line) and the filtered one (thick line). The $S/N$, taking the Fe\,{\sc i}
line at 630.2 nm as reference, $S/N$ is 3.358. Note that the recovered Stokes $V$
profiles are almost overlapping the original ones.}
\label{sn3360}
\end{figure}

\subsection{Intermediate $S/N$}
As representative of an intermediate $S/N$, we present the results
obtained when the $S/N$ in the 630.2 nm line is 3.358. As can be seen in Fig. \ref{ampl_v} 
amplitudes like the one of the $630.2$ nm are not very common in the spectrum. This means that 
most of the spectral lines would have S/N values at least 5 to 10 times smaller. Then, we are dealing with 
a an example that can be representative
of a typical observational case in stellar polarised spectra.

Although the real signal is still below the
noise level for most of the lines, the number of selected eigenvectors is 
4 according to the criterium of Sec. \ref{sec:procedure}. 
The left panels of Fig. \ref{sn3360} show the comparison between
the original synthetic profile without noise and the profile recovered after PCA
denoising starting from the noisy profiles. The right panels of Fig.
\ref{sn3360} show the comparison between the noisy and the PCA-filtered signals
of three individual Fe\,{\sc i} spectral lines and a Cr\,{\sc i} line widely known in solar 
physics. In these conditions, the shape of all the spectral lines is roughly reproduced and 
the $S/N$ of the filtered data is good enough for a reliable study of these individual spectral lines. 
The results presented so far have been obtained using the automatic criterium for the selection of eigenvectors to be
included into the linear combination. This facilitates the statistical analysis of the method
and gives an idea of how the methods behave with real data. However, we want to point out that 
better results are obtained in particular cases when one
carefully selects the eigenvectors used to reconstruct the signal. An example of this is the lowest
panel of Fig. \ref{sn3360}, where the filtered signal has a spurious contribution from high-order
PCA eigenvectors that can be improved by taking less PCA eigenvectors in the reconstruction.

\begin{figure}[!t]
\includegraphics[width=0.49\columnwidth]{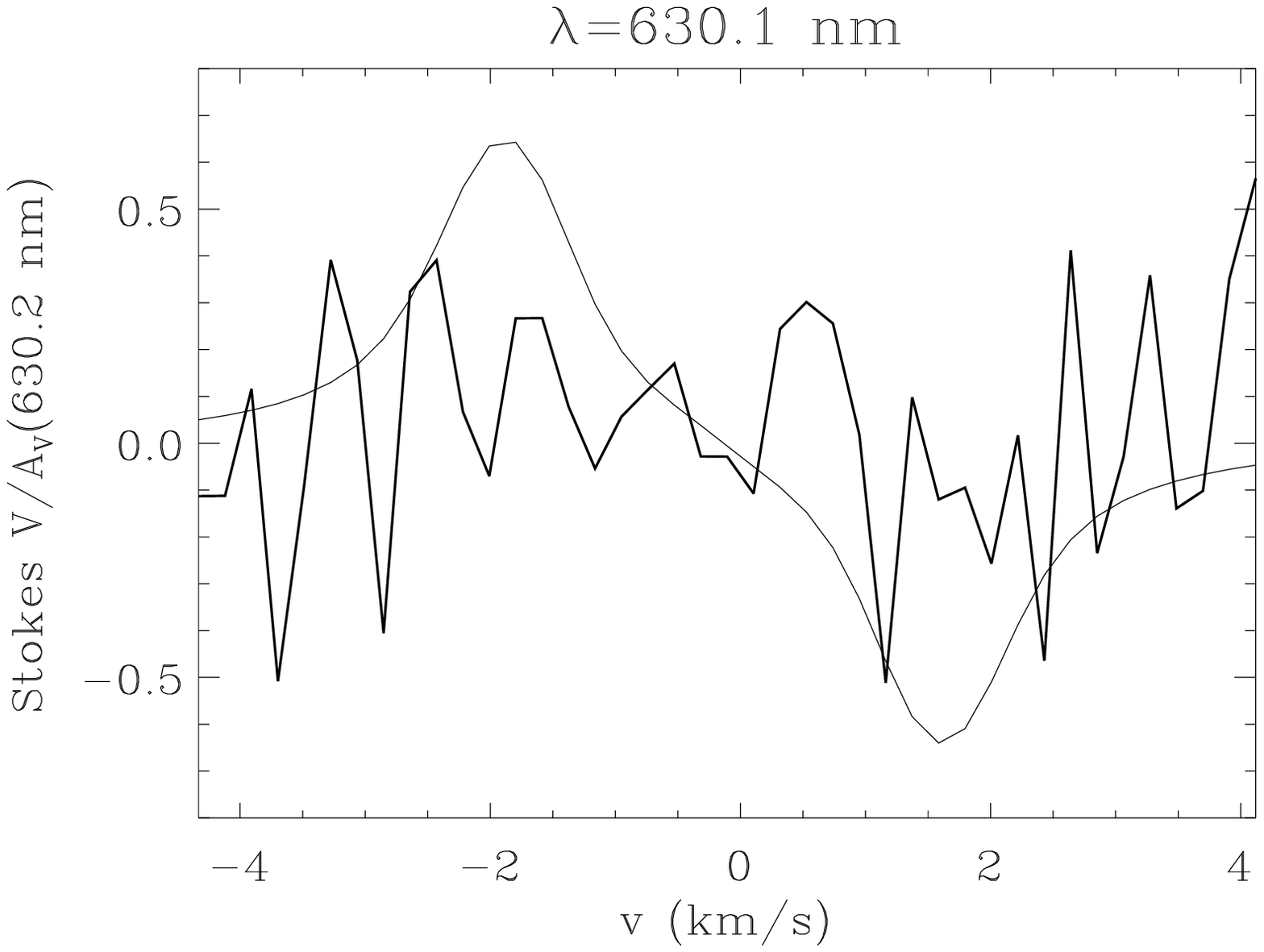}
\includegraphics[width=0.49\columnwidth]{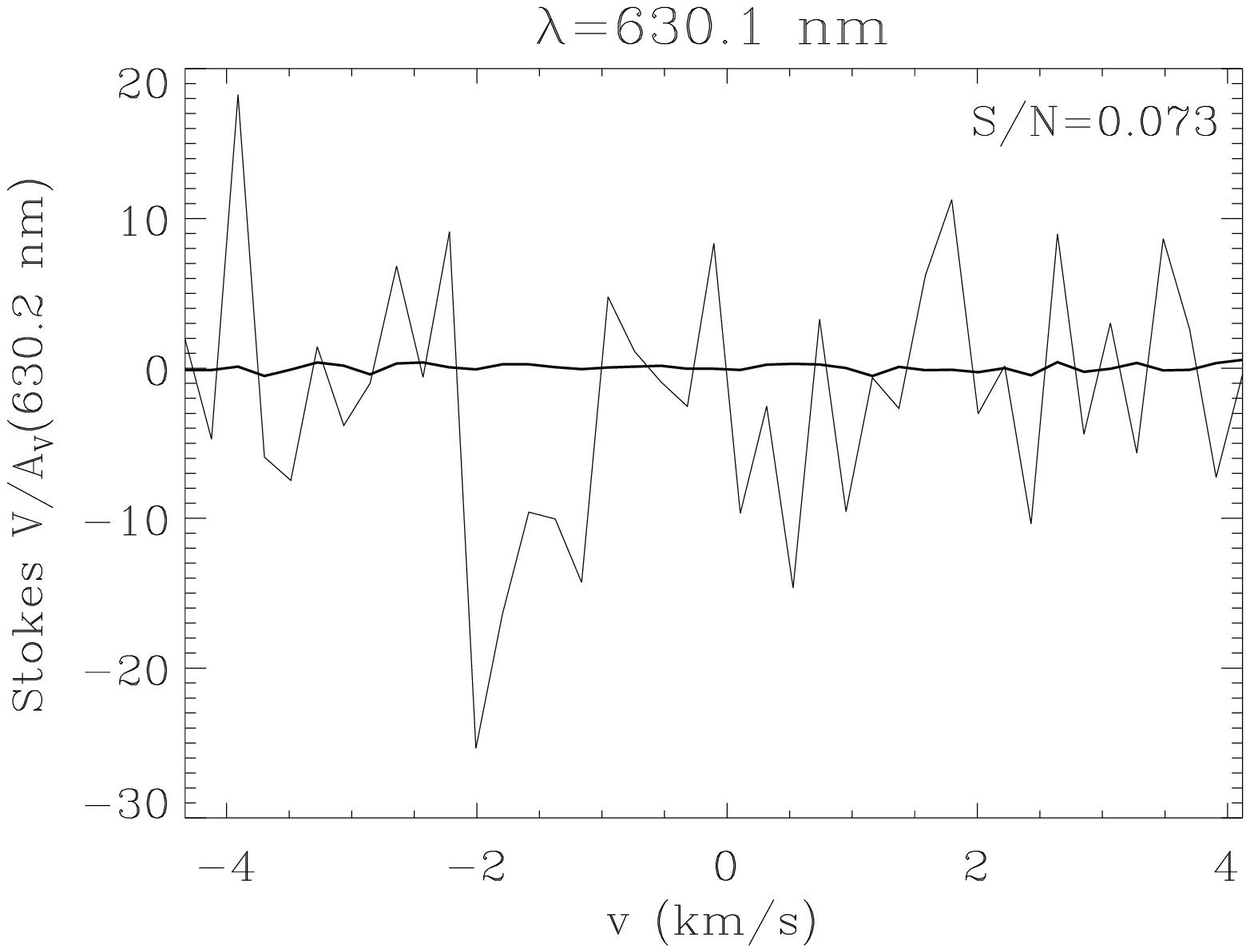}
\includegraphics[width=0.49\columnwidth]{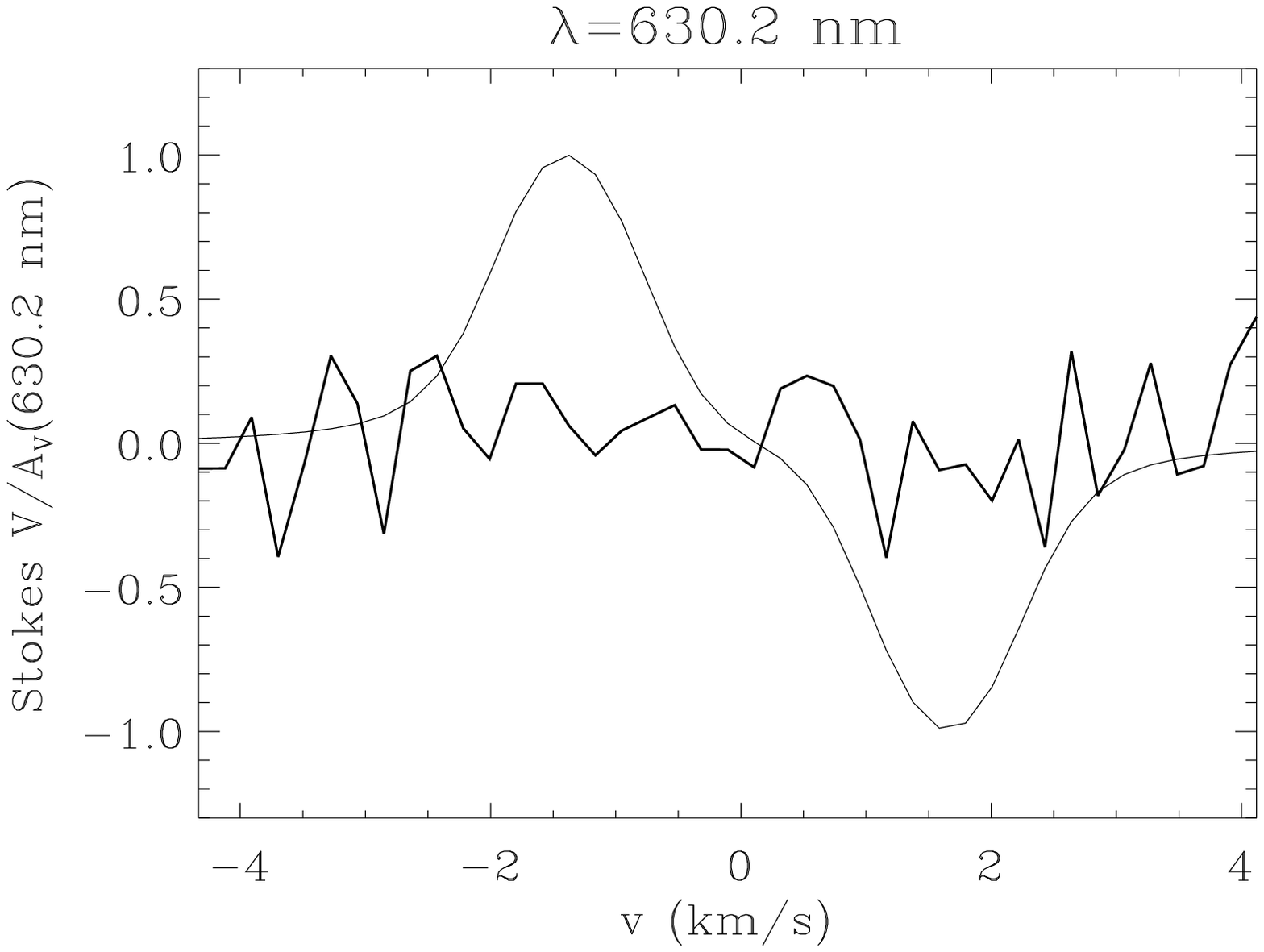}
\includegraphics[width=0.49\columnwidth]{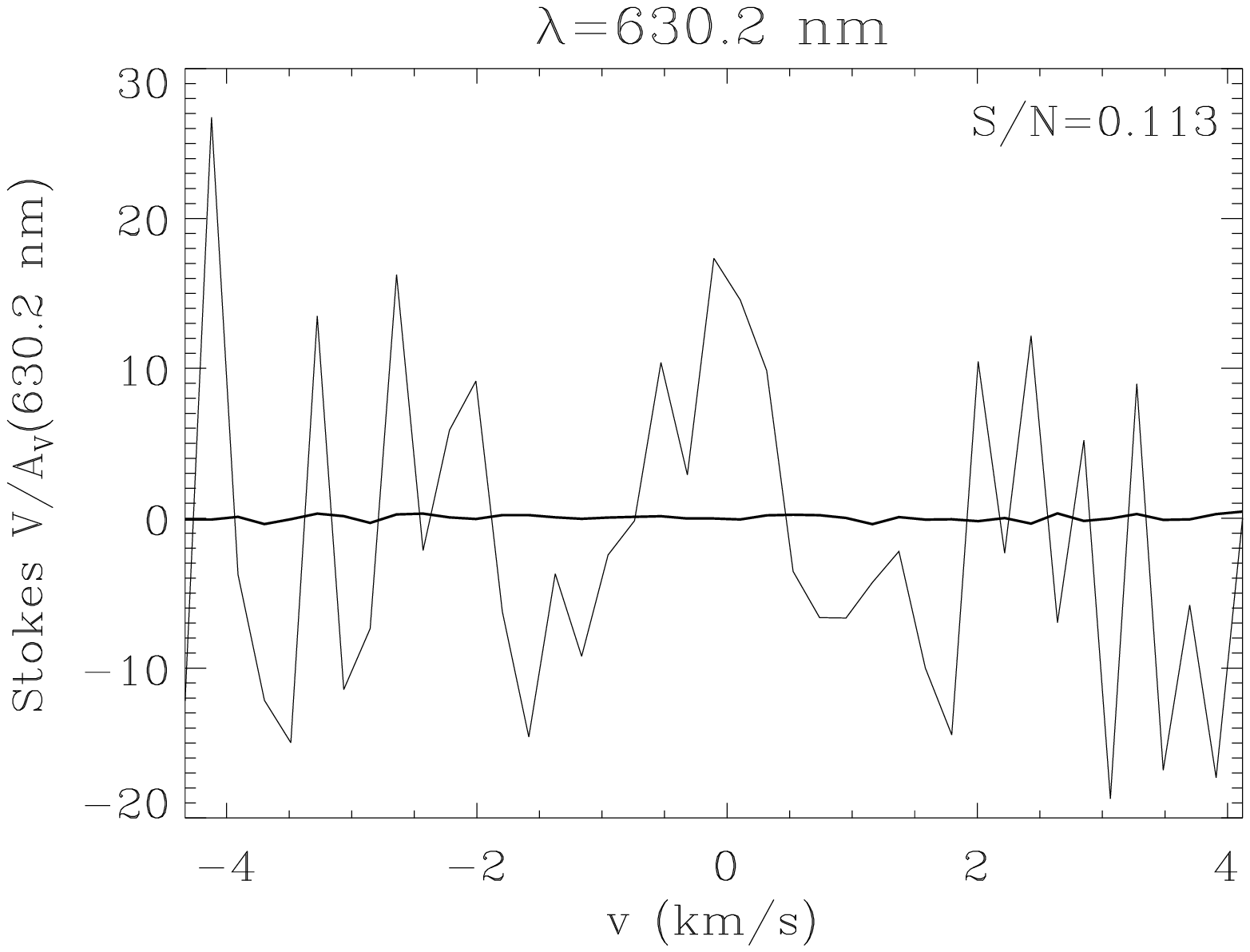}
\includegraphics[width=0.49\columnwidth]{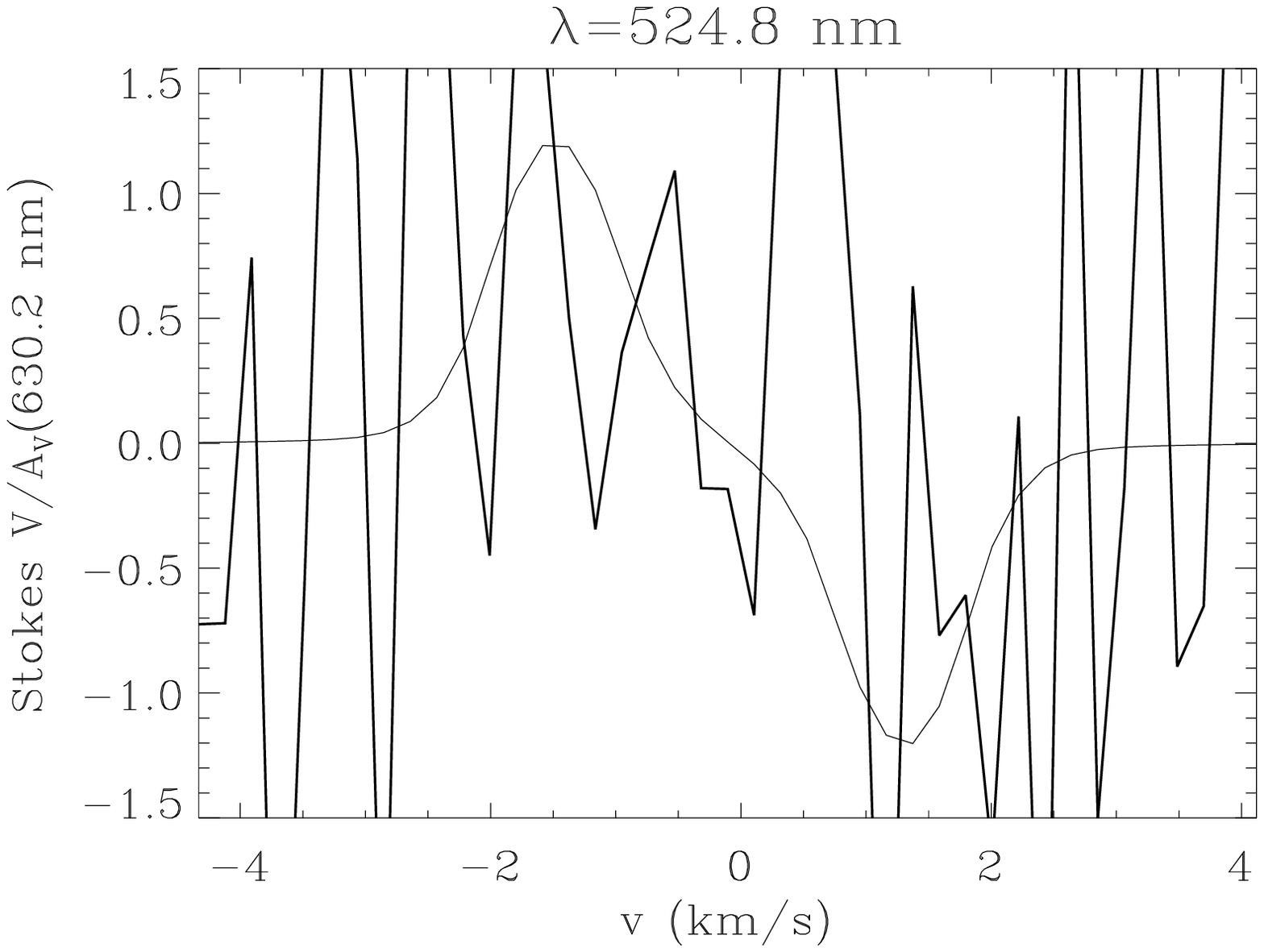}
\includegraphics[width=0.49\columnwidth]{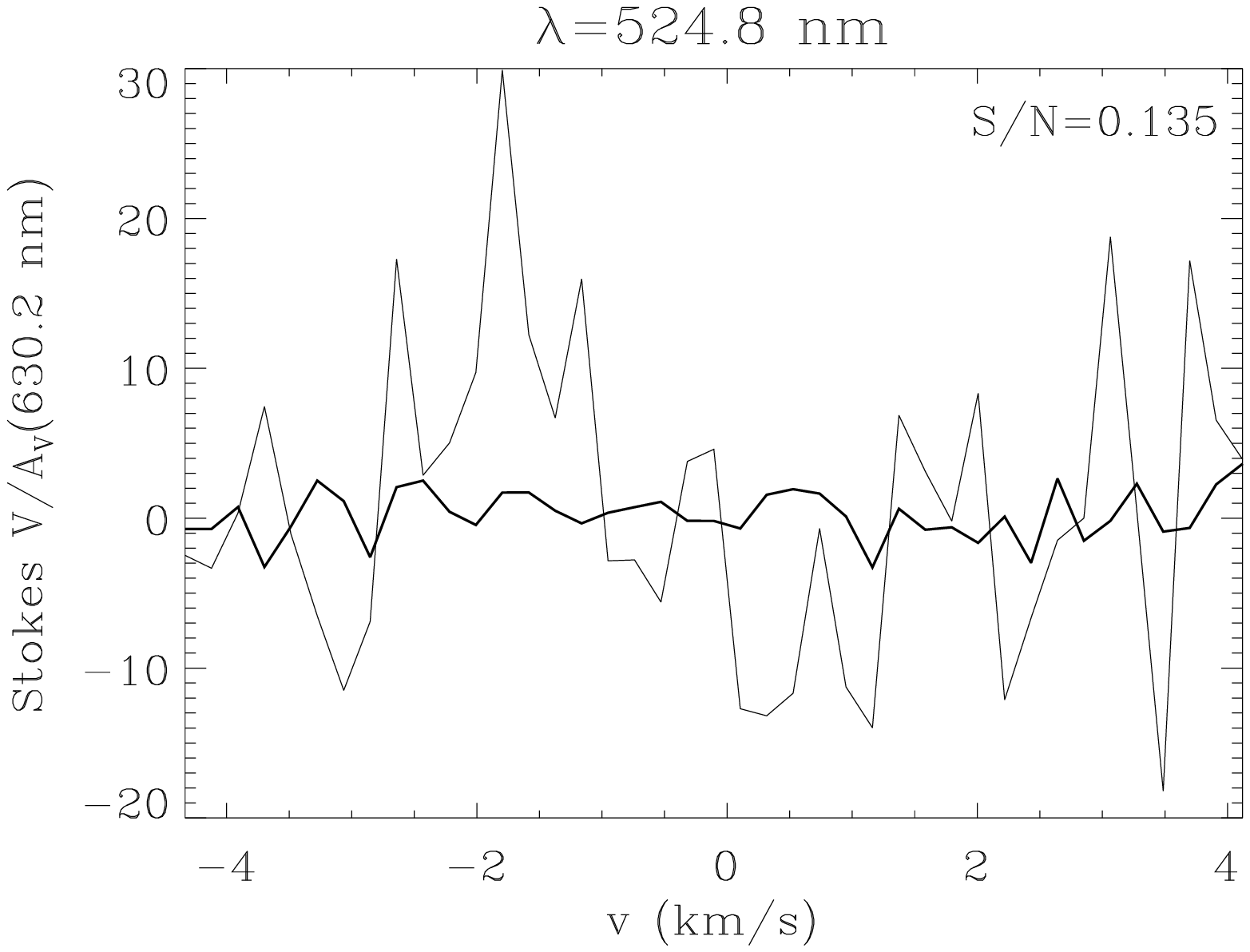}
\includegraphics[width=0.49\columnwidth]{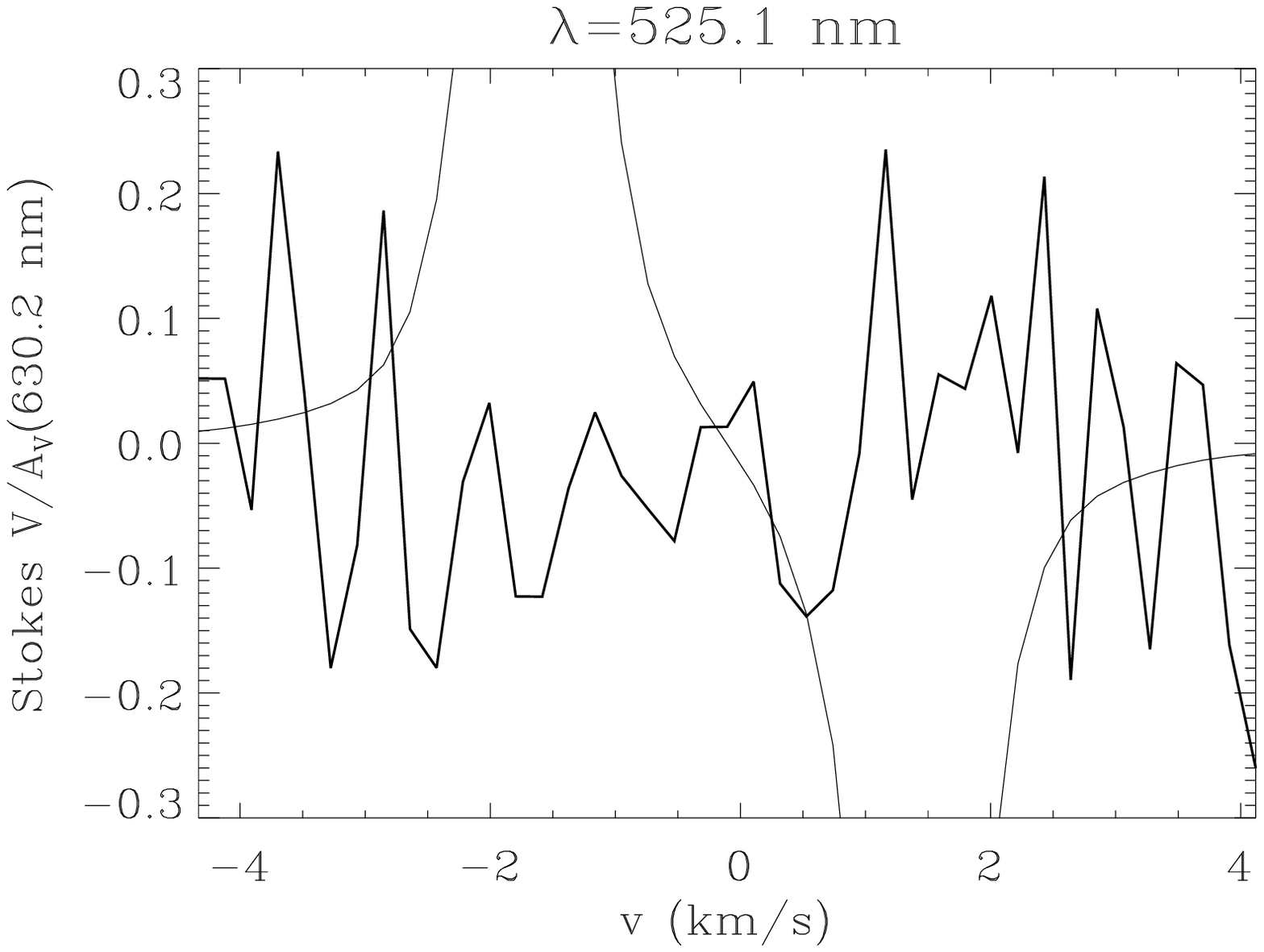}
\includegraphics[width=0.49\columnwidth]{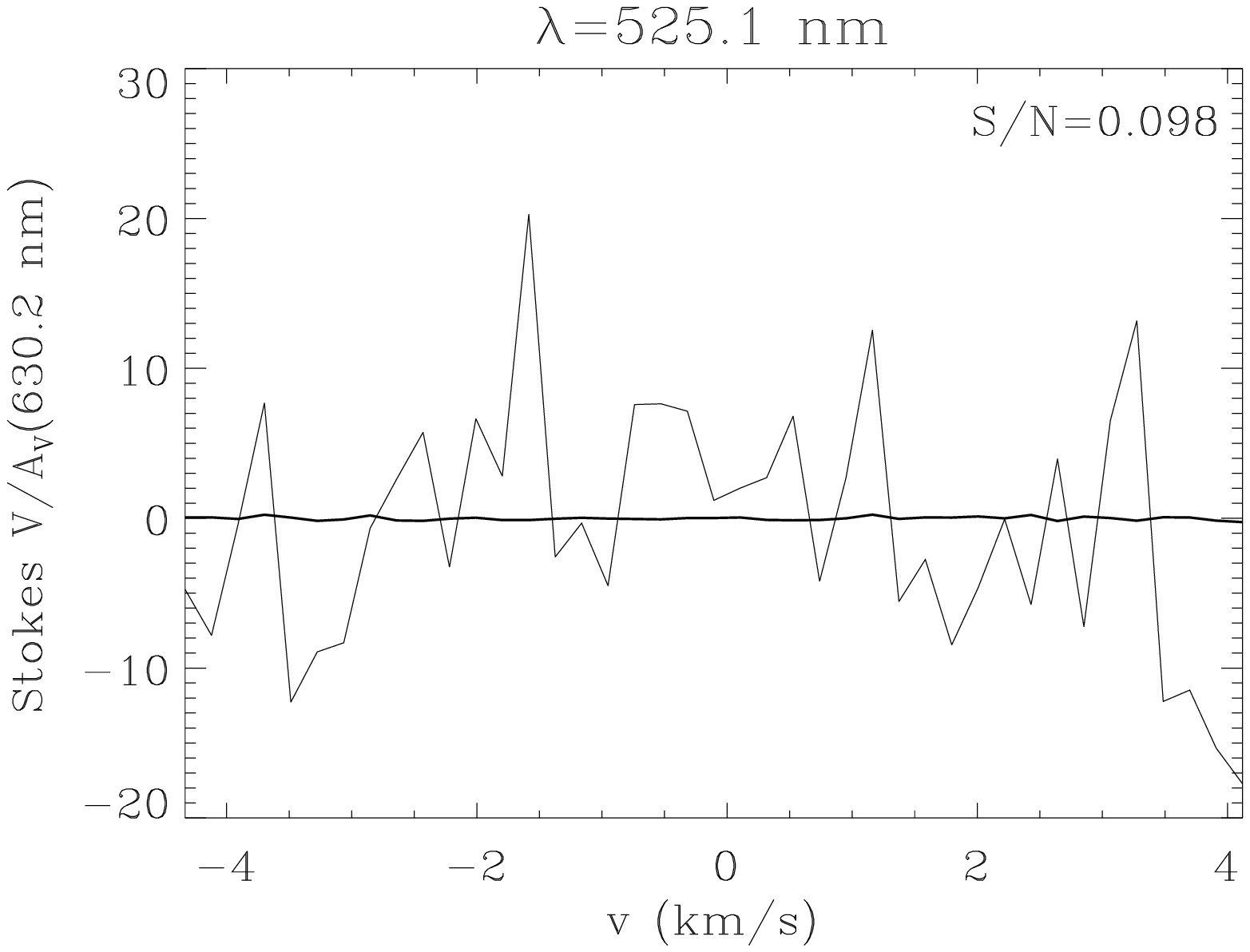}
\caption{The same as in Fig. \ref{sn3360} but for $S/N$ is 0.113.}
\label{sn0113}
\end{figure}

Note that the results shown in Figs. \ref{sn3360}, \ref{sn0785} and \ref{sn0113} correspond to a particular
noise realization. The first eigenvector and the projection of the data onto it
do change for different noise realizations (note also that the sign of the filtered
profile can, in some cases, be the opposite to the original one). This will be explained in 
Section \ref{sec:trends}.

\subsection{Low $S/N$}
\begin{figure}[!t]
\includegraphics[width=0.49\columnwidth]{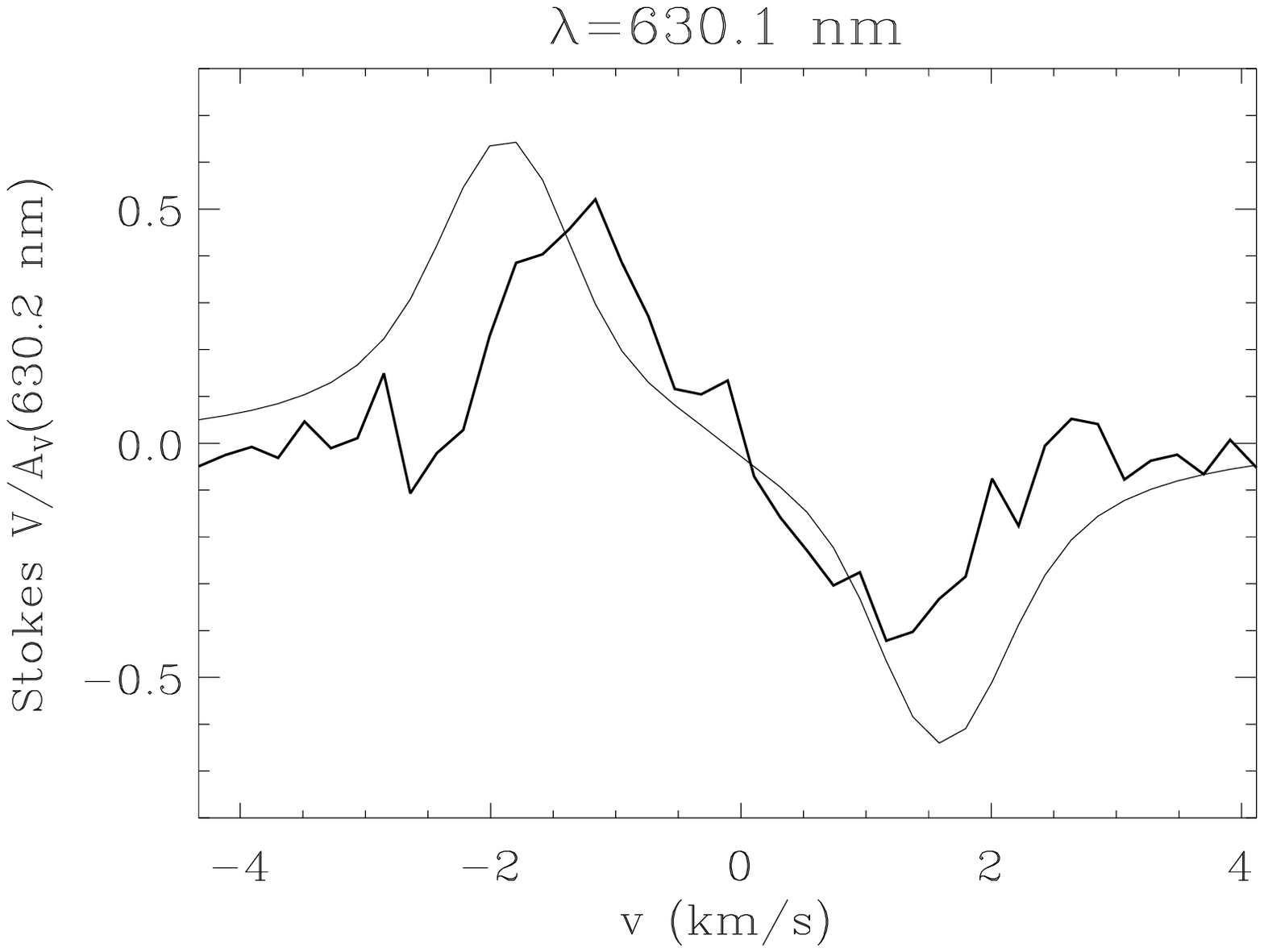}
\includegraphics[width=0.49\columnwidth]{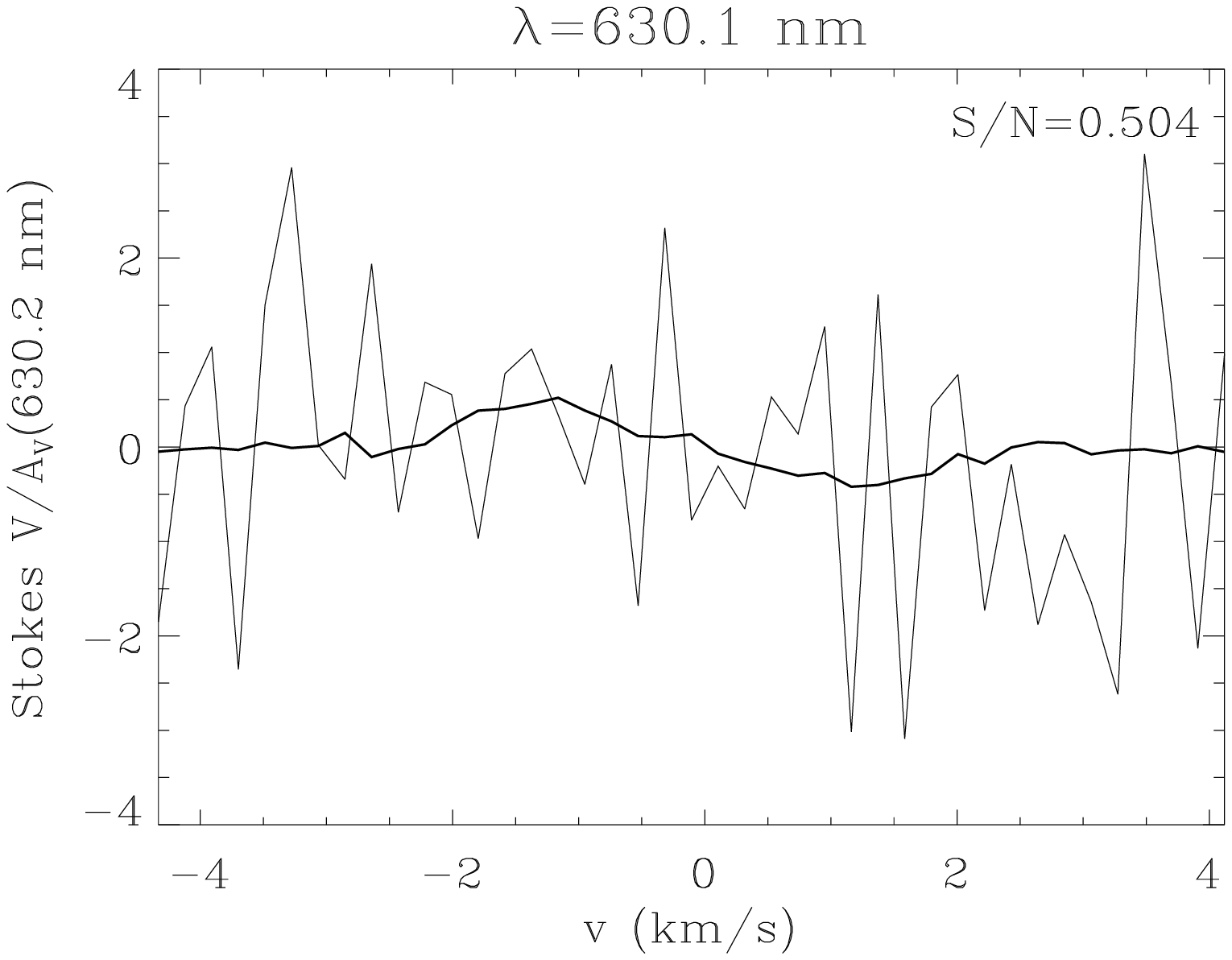}\\
\includegraphics[width=0.49\columnwidth]{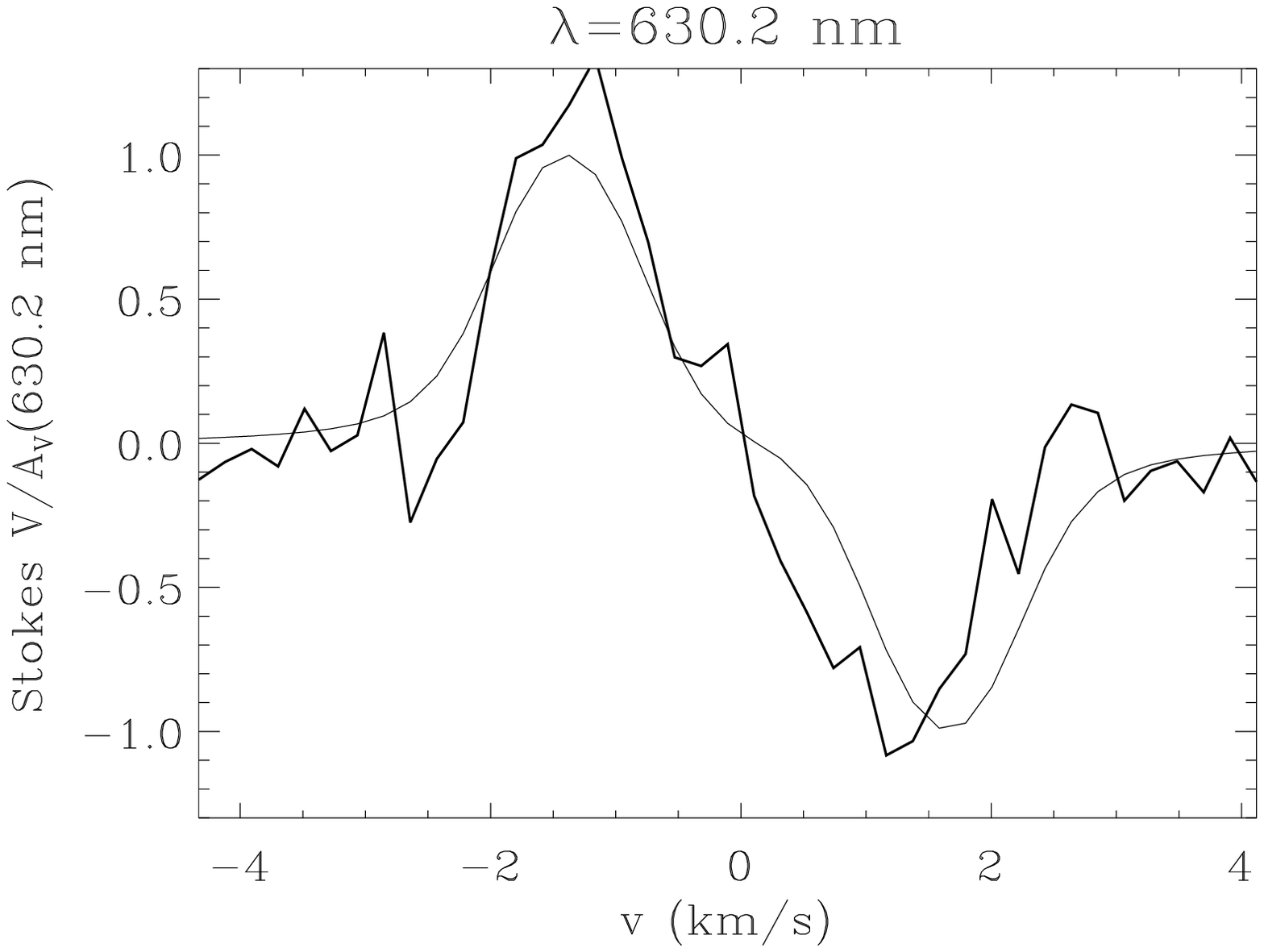}
\includegraphics[width=0.49\columnwidth]{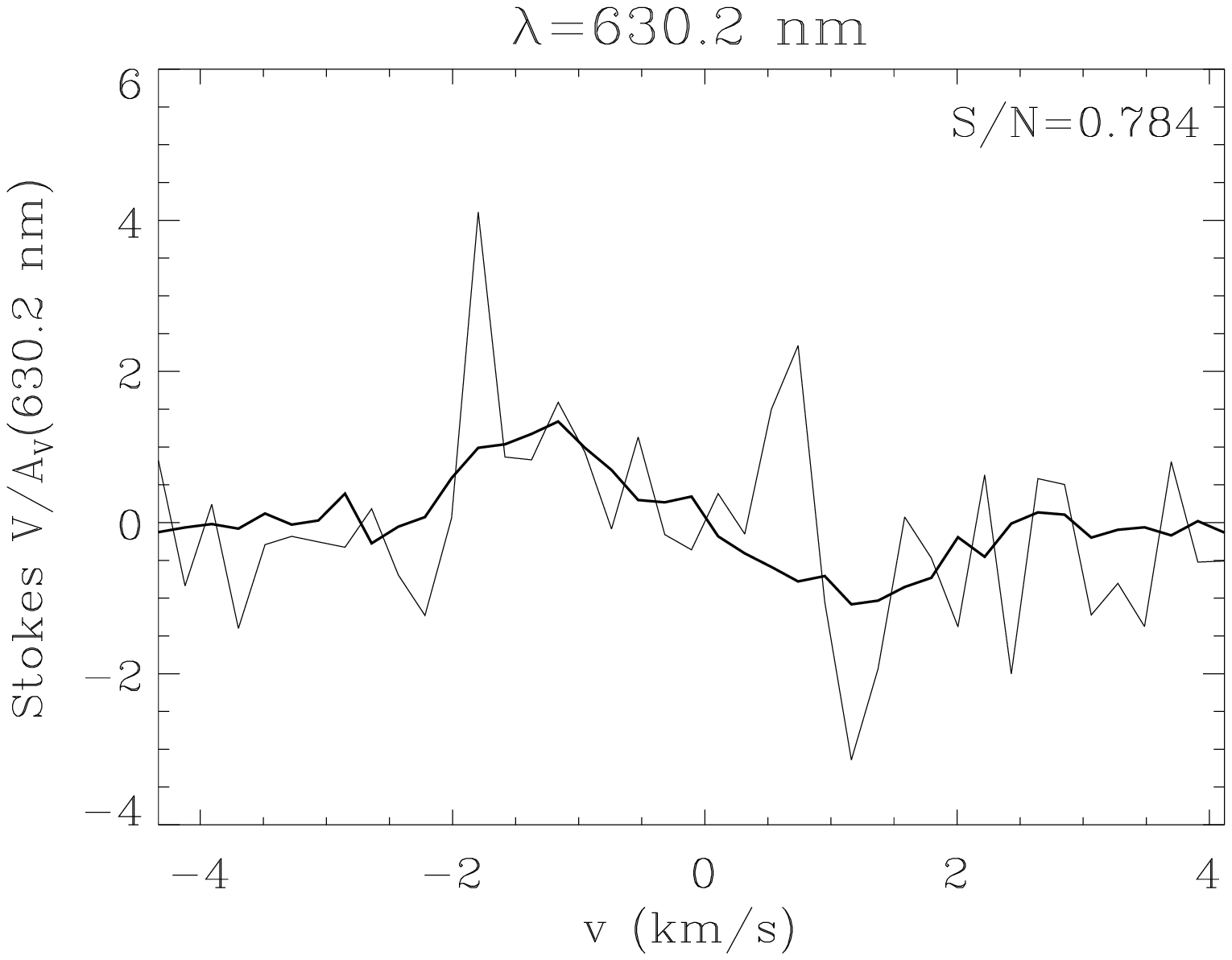}\\
\includegraphics[width=0.49\columnwidth]{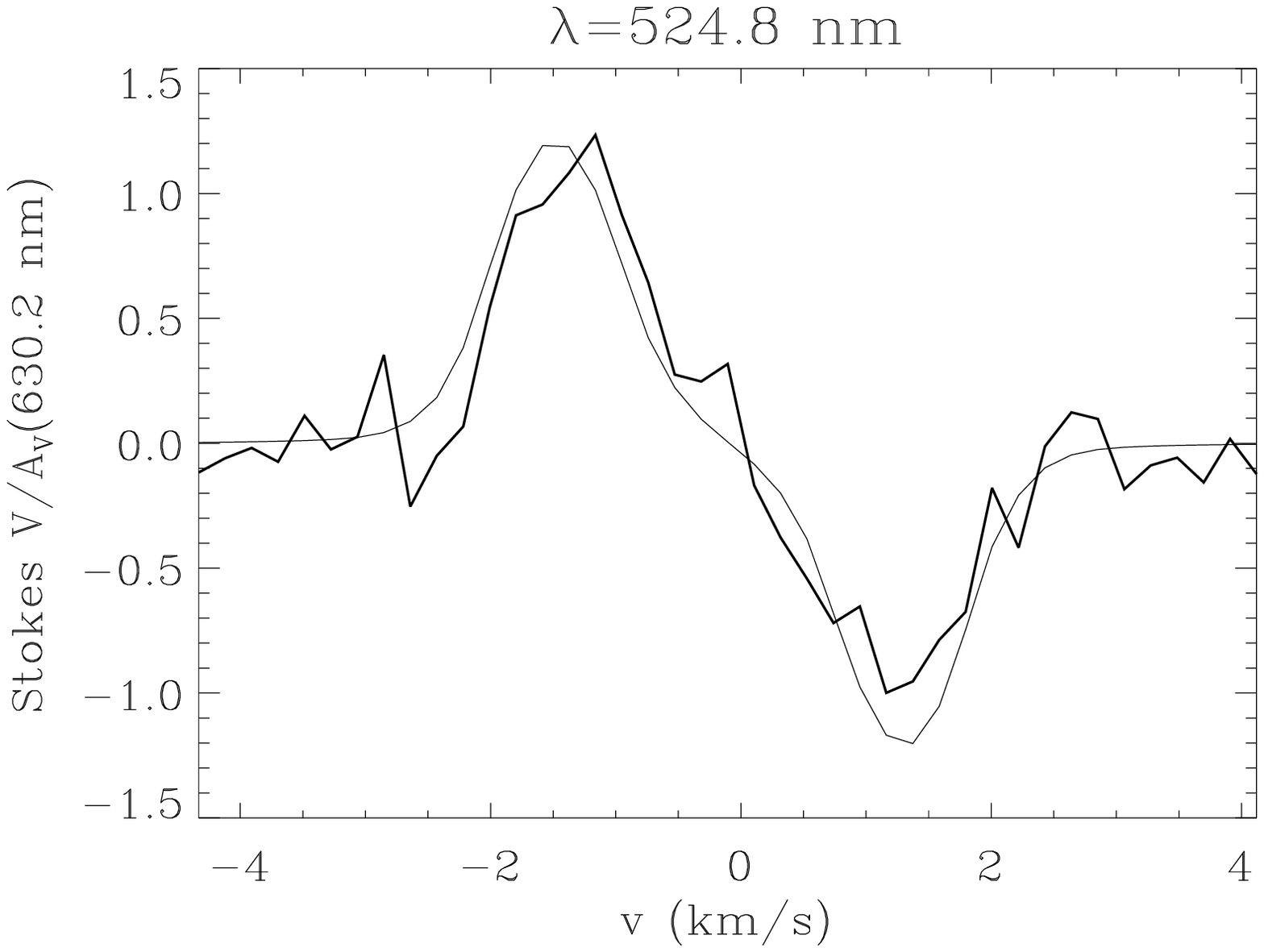}
\includegraphics[width=0.49\columnwidth]{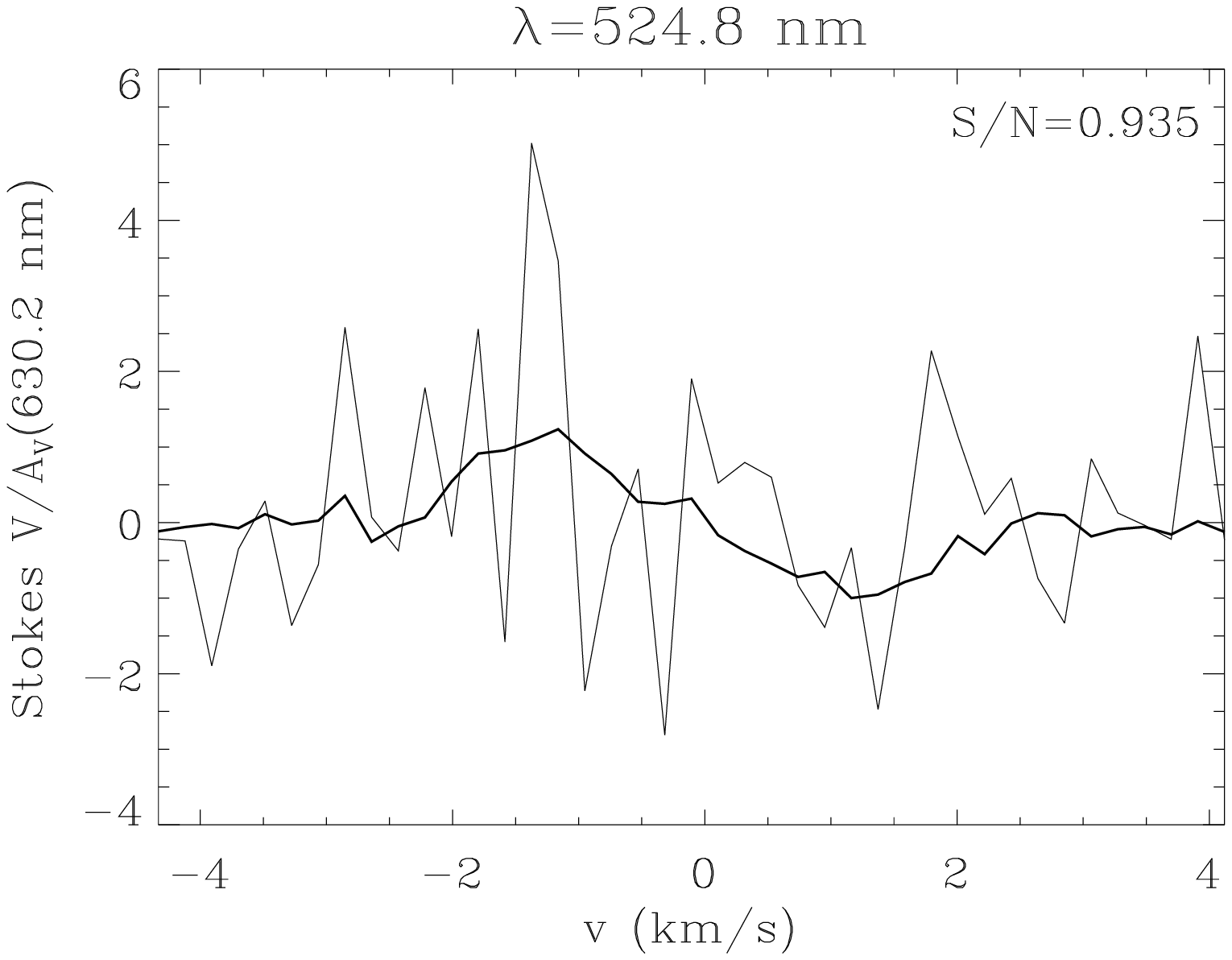}\\
\includegraphics[width=0.49\columnwidth]{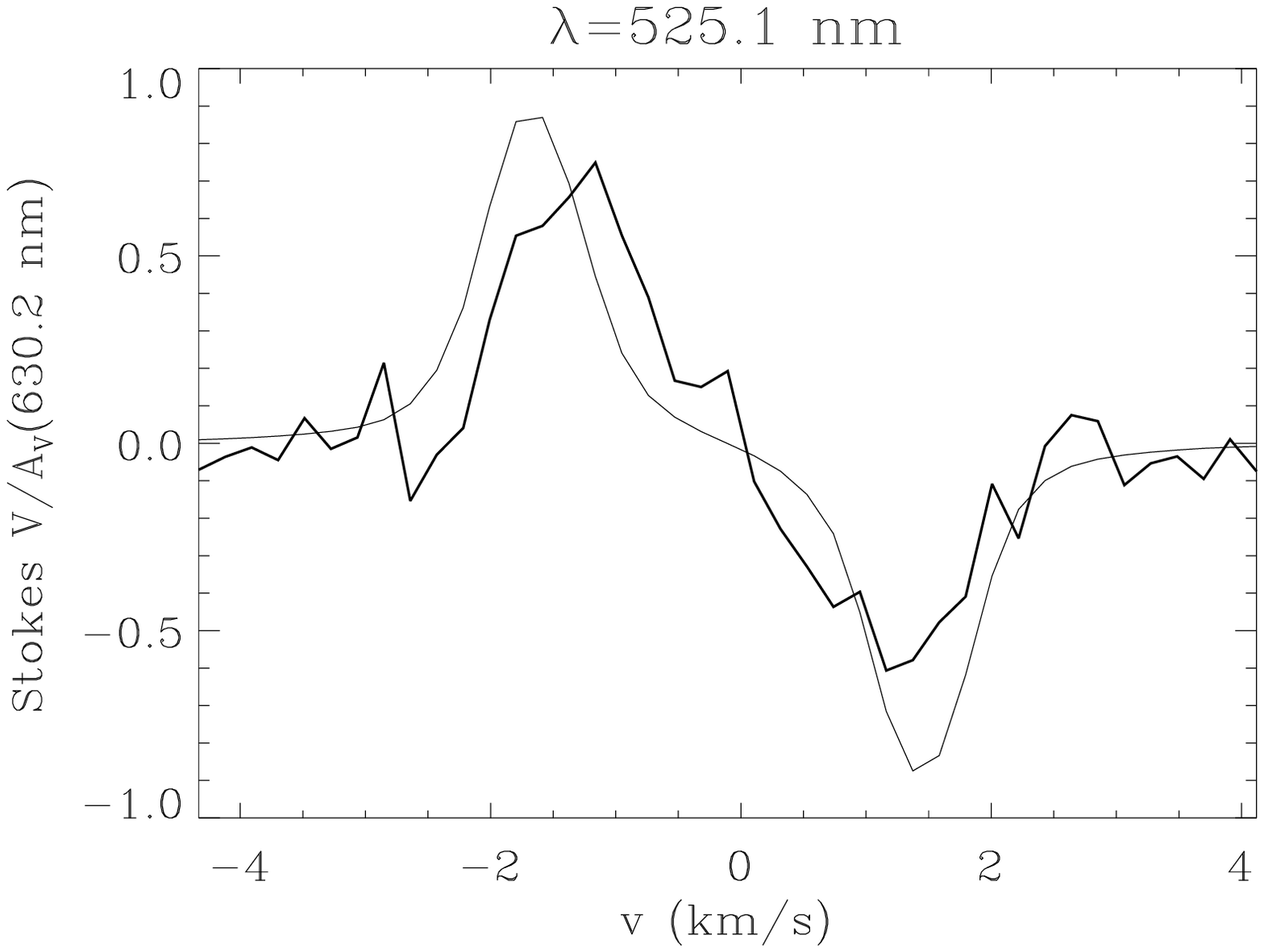}
\includegraphics[width=0.49\columnwidth]{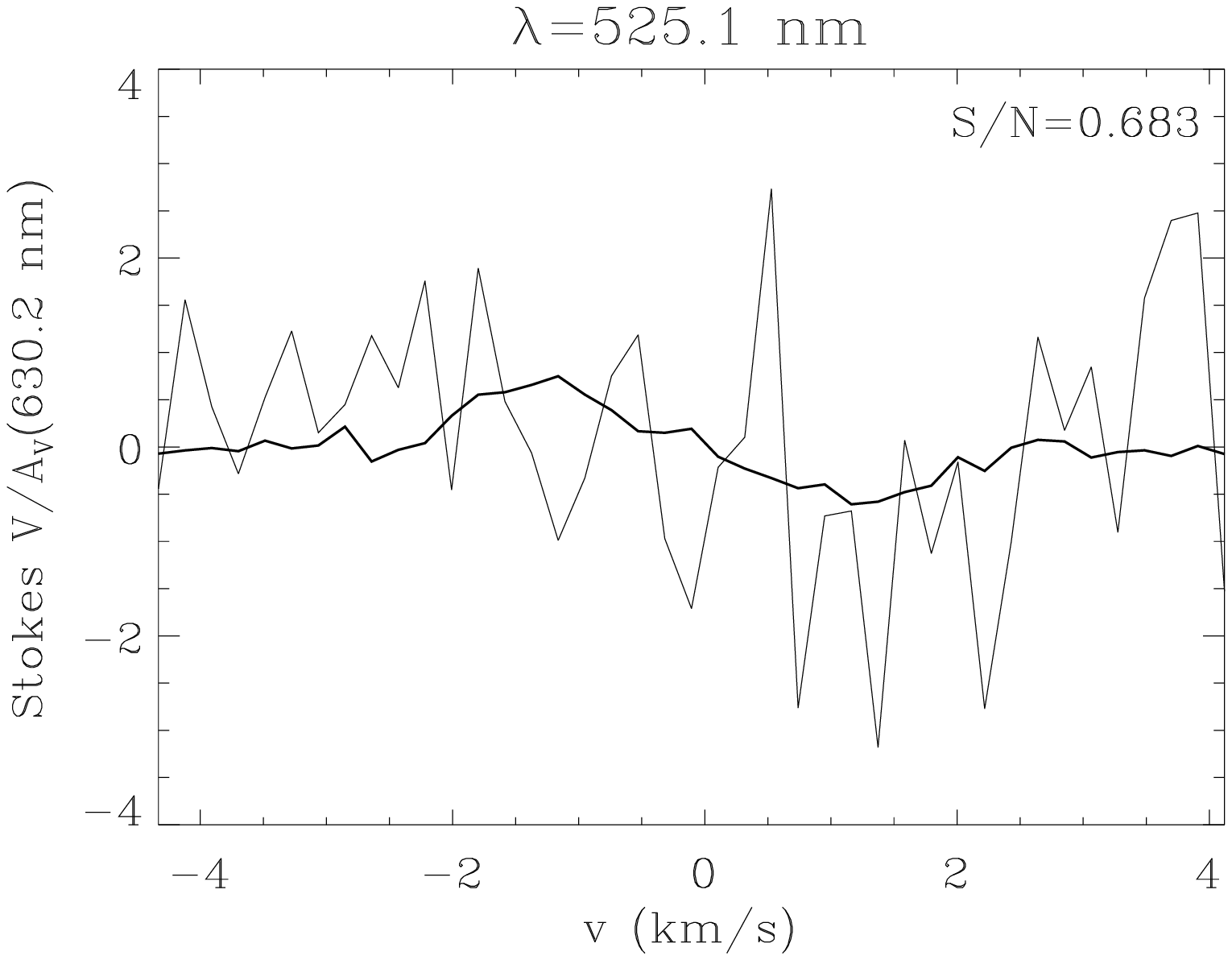}
\caption{The same as in Fig. \ref{sn3360} but for $S/N$ is 0.784.}
\label{sn0785}
\end{figure}

We present in this section the results when the $S/N$ of the \hbox{630.2 nm} 
spectral line has been decreased to 0.784. Figure \ref{sn0785} shows the comparison between the original and the 
filtered signals (left panels) and between the noisy and filtered observations for the 
same four spectral lines. According to the criterium presented in Sec. \ref{sec:procedure},
we reconstruct the data taking into account only the first eigenvector. 
This case is even worst than the typical scenario we can expect from real
spectro-polarimetric observations \citep[e.g.,][]{donati97}. Fig. \ref{sn0785}
shows that the $S/N$ of the filtered data has been considerably improved. The shapes of the spectral
lines can now be foreseen under the noise, mainly in the first and third panels. However, note that
since we are taking into account only the first
eigenvector, many details are still not reproduced. This means that all spectral lines should have the same 
shape, the only difference between them being the projection coefficient. 
However, the improvement in $S/N$
has allowed to unambiguously detect the presence of circular polarization signals in some lines.

\subsection{Very low $S/N$}
As a representative of a case in which the signal is far below the noise level,
we have chosen a noise distribution with a $S/N$ in the 630.2 nm line of 0.113. The signal to noise ratio in each individual line is indicated in each panel. 
Only the first eigenvector has been used for reconstructing the data set
according to the criterium described above.  It is evident that the
signal has been strongly filtered and that the noise level is much lower than in the
simulated observations. In this case, the value of $S/N$ is so 
small that the information about the line profiles can not be extracted from the noisy
data.

\subsection{Denoising trends in $S/N$}
\label{sec:trends}
\begin{figure*}[!ht]
\includegraphics[width=\columnwidth]{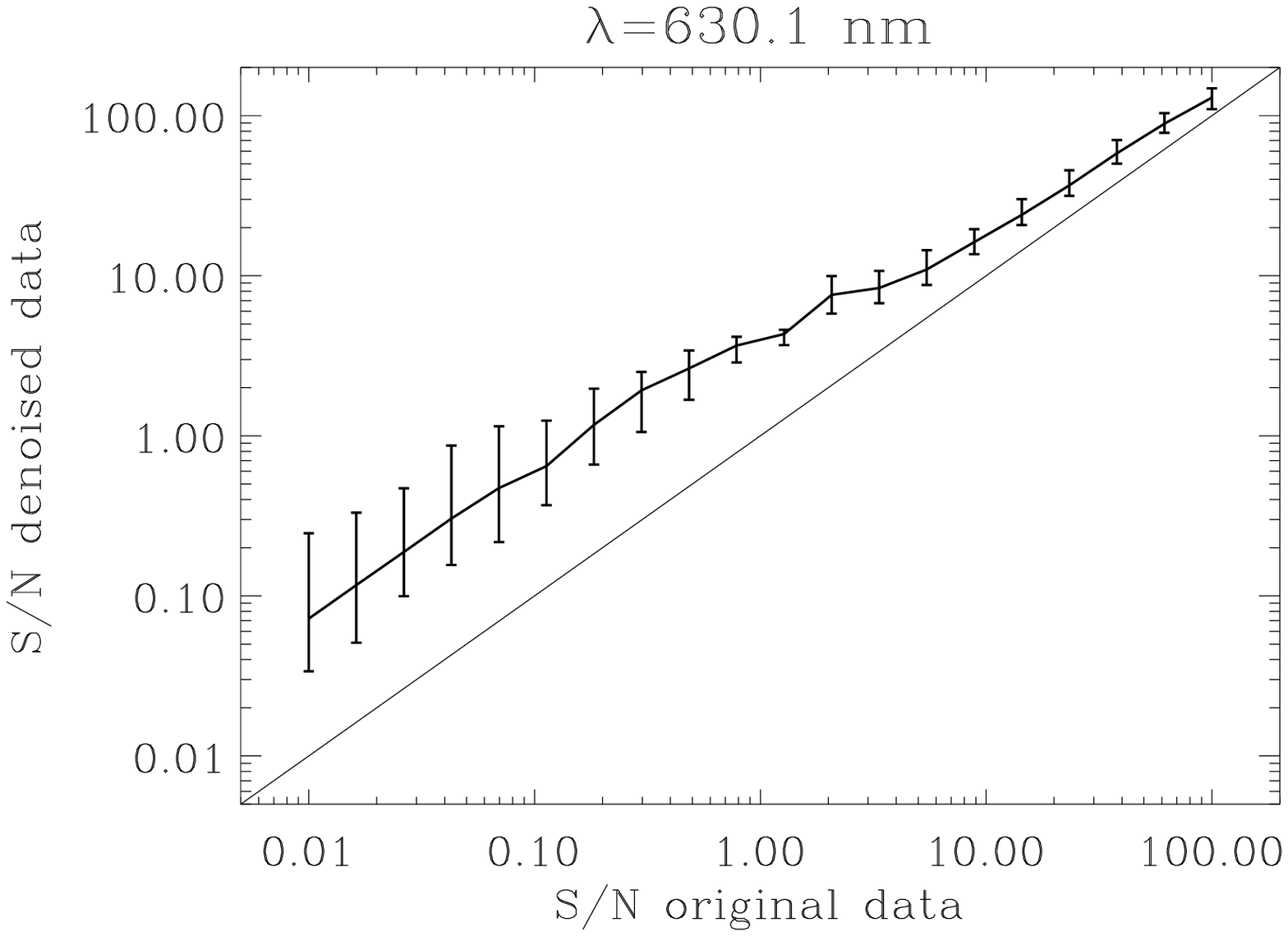}
\includegraphics[width=\columnwidth]{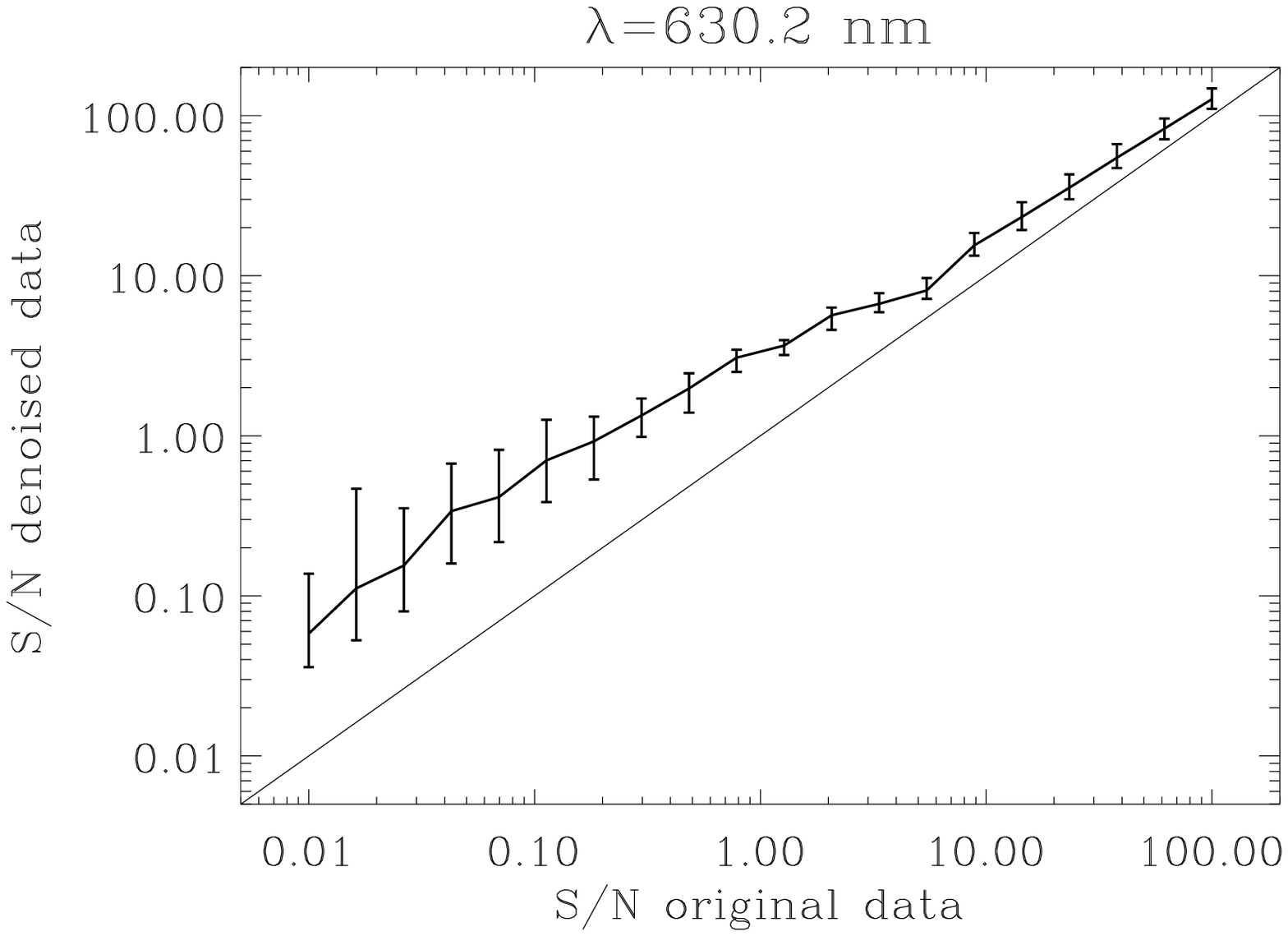}\\
\includegraphics[width=\columnwidth]{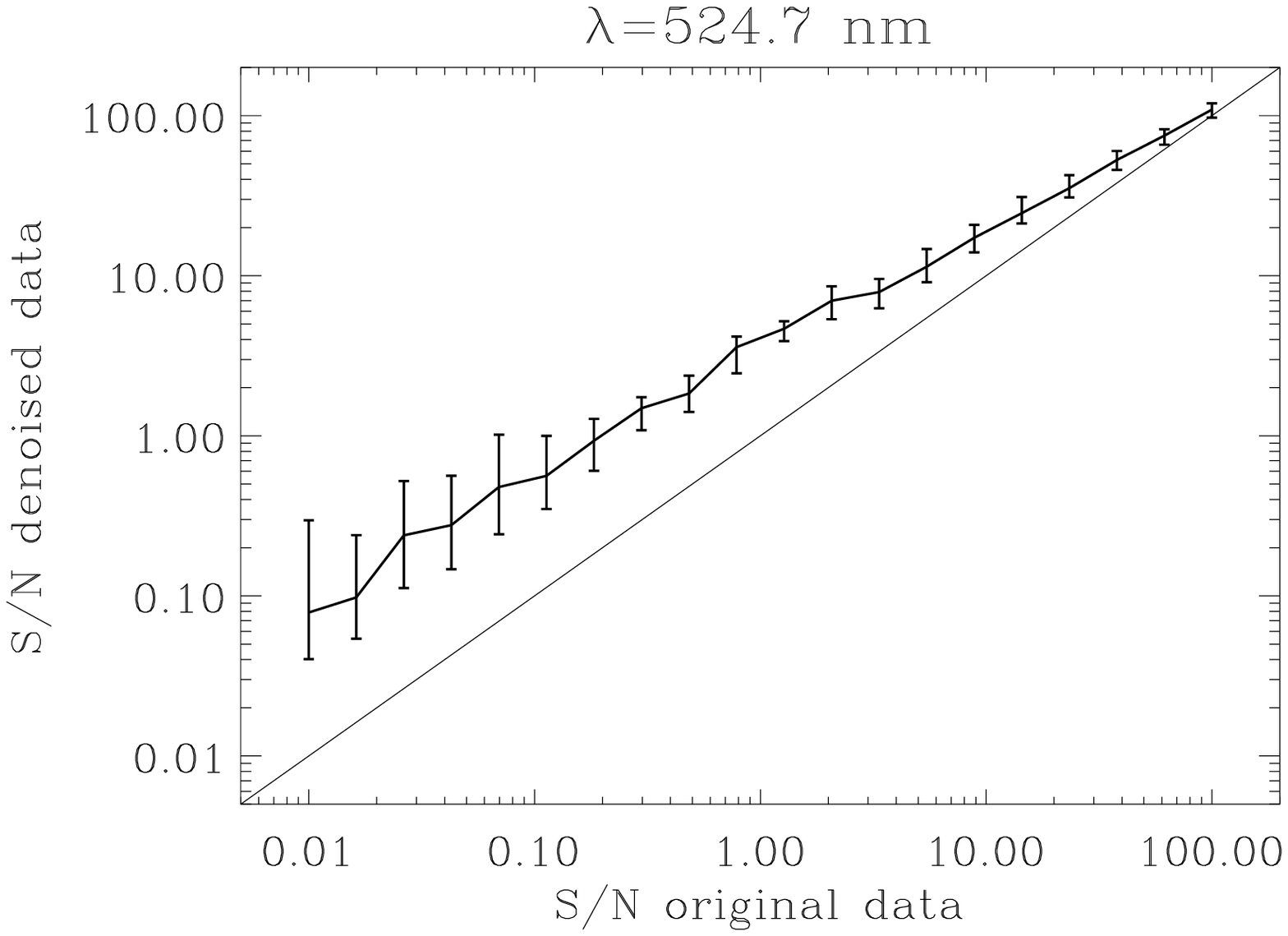}
\includegraphics[width=\columnwidth]{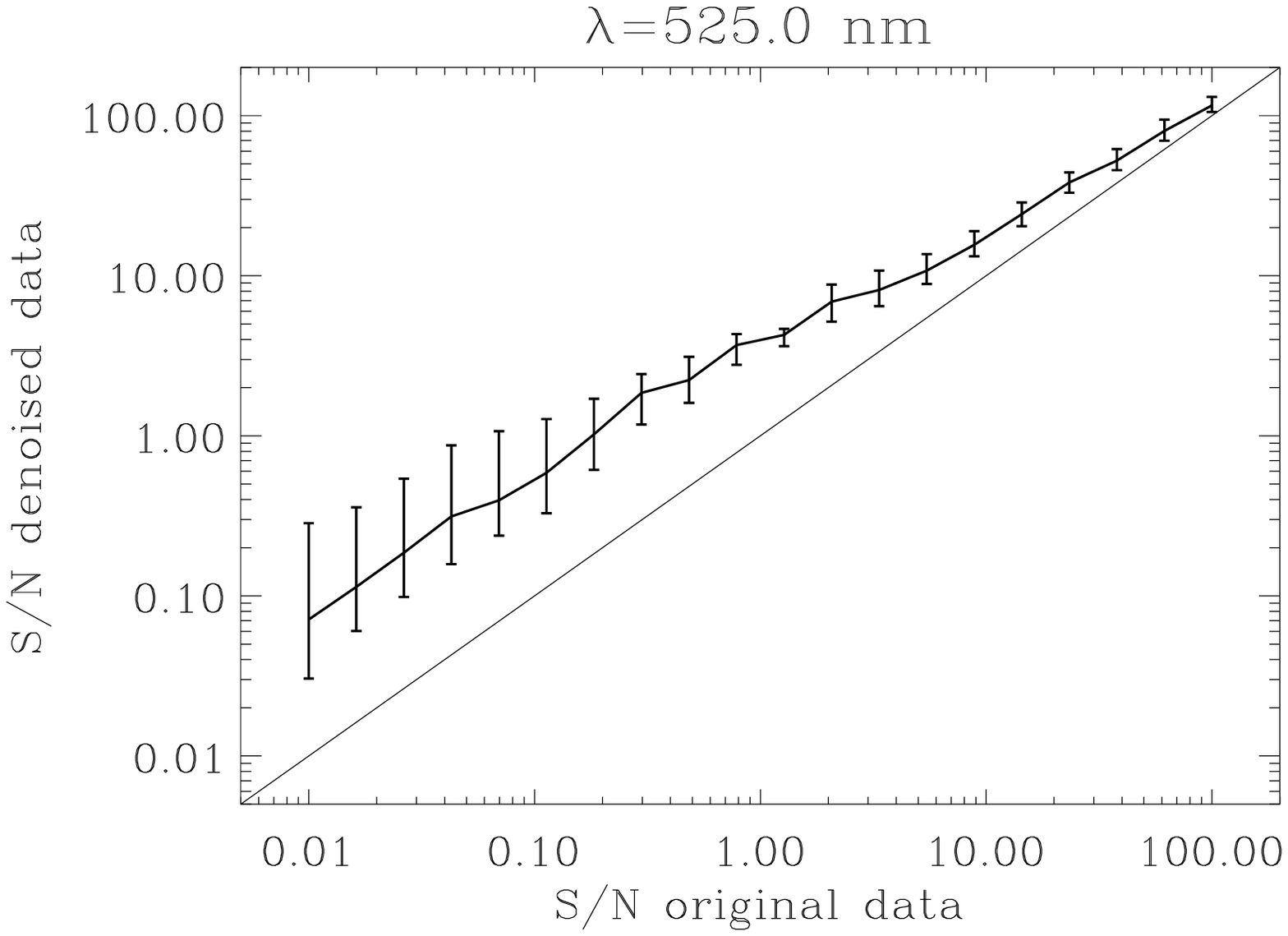}
\caption{Signal to noise ratio of the filtered data versus the original $S/N$ of
the observations for Stokes $V$. The error bars indicate the 68\% confidence interval obtained in
a Montecarlo procedure with different noise realizations.
}
\label{final}
\end{figure*}

\begin{figure*}[!t]
\includegraphics[width=\columnwidth]{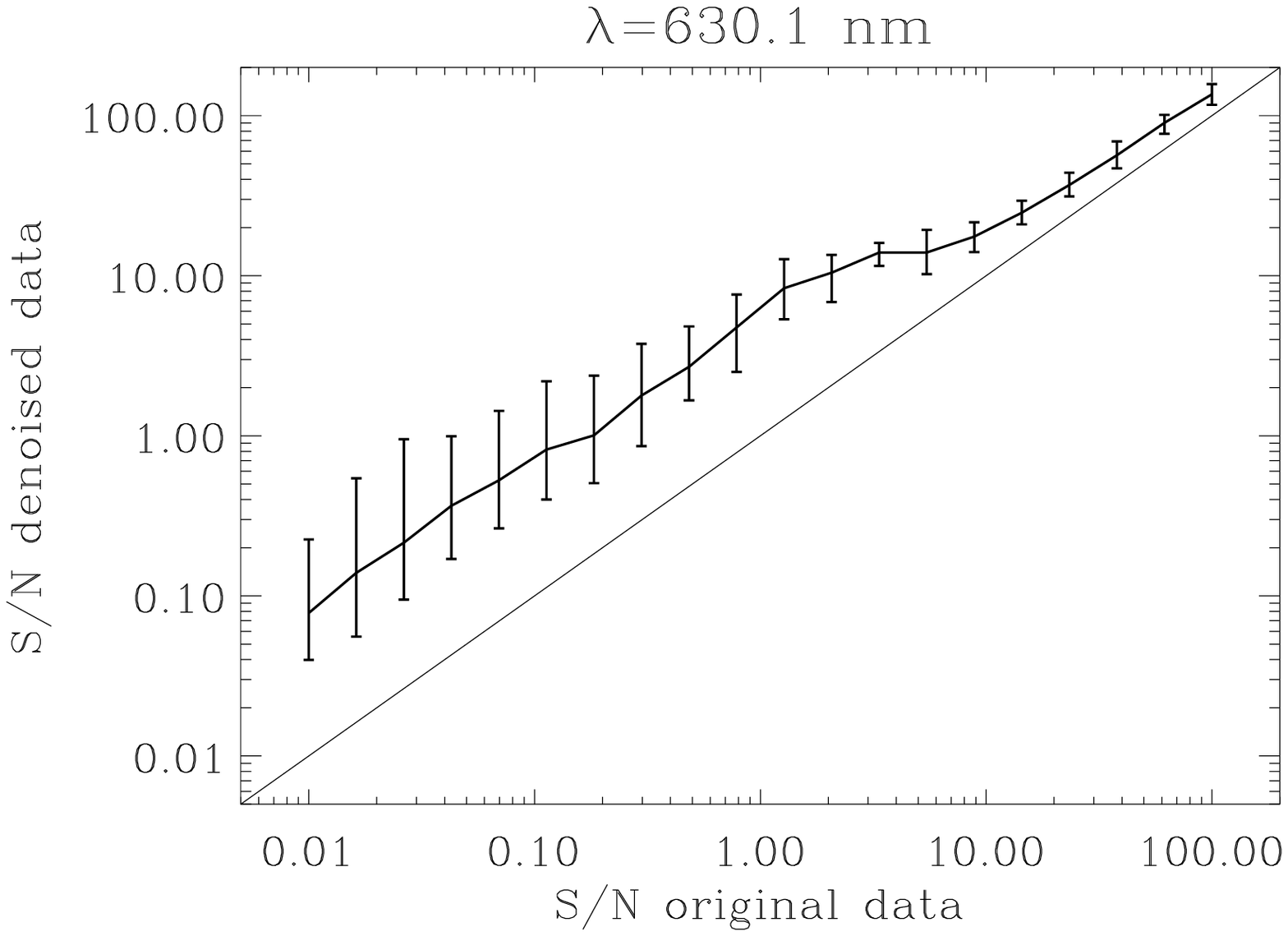}
\includegraphics[width=\columnwidth]{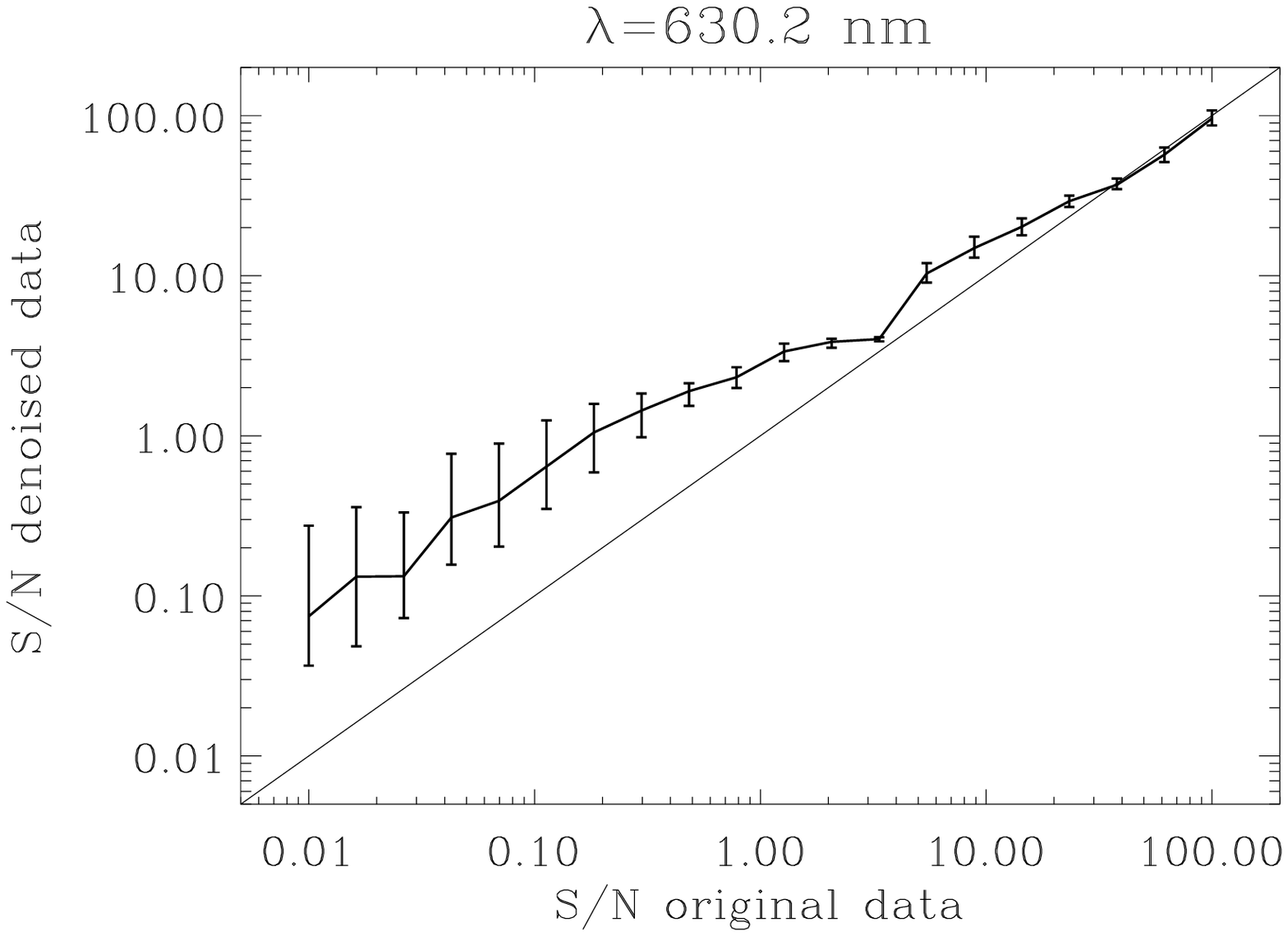}\\
\includegraphics[width=\columnwidth]{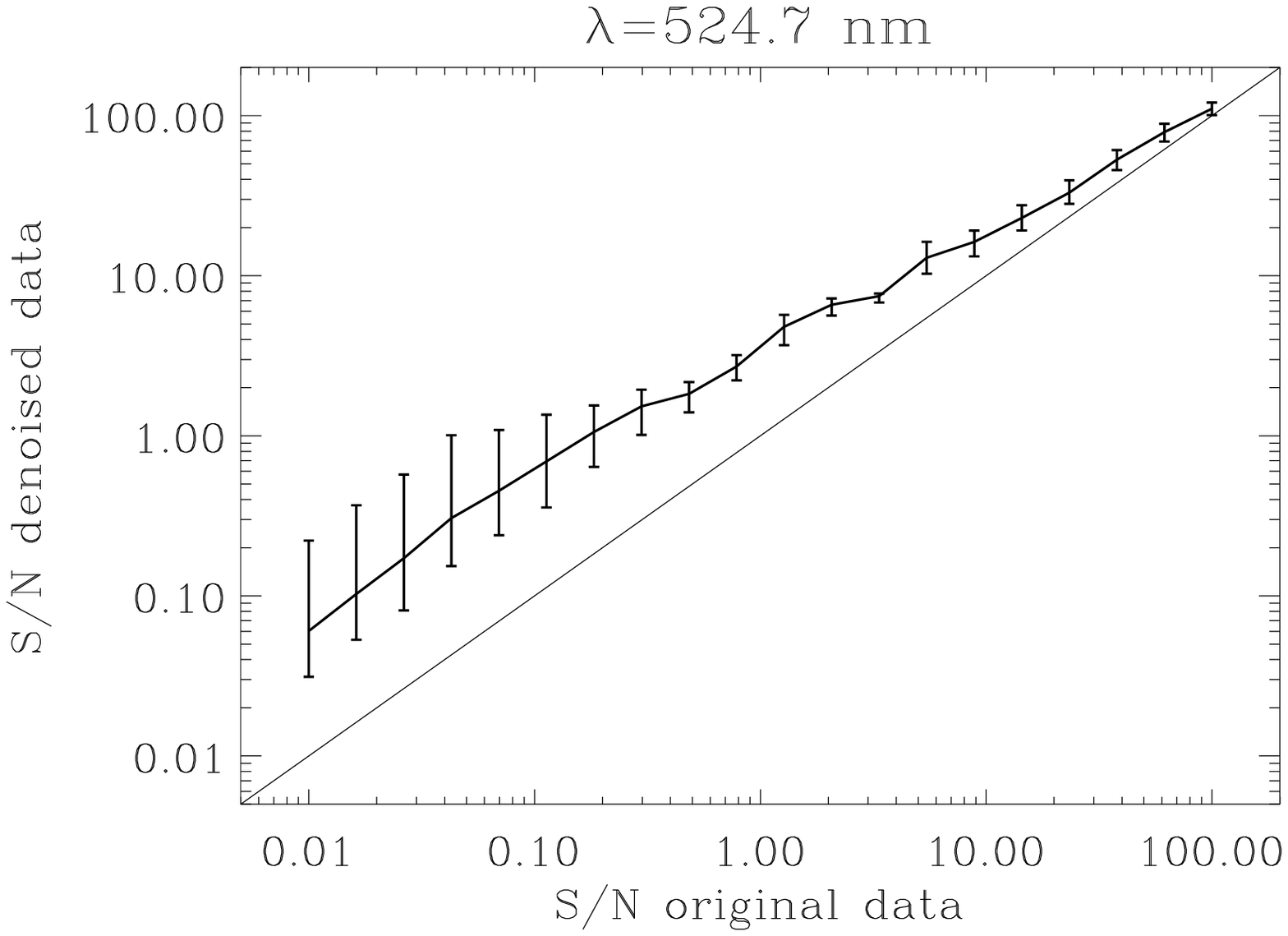}
\includegraphics[width=\columnwidth]{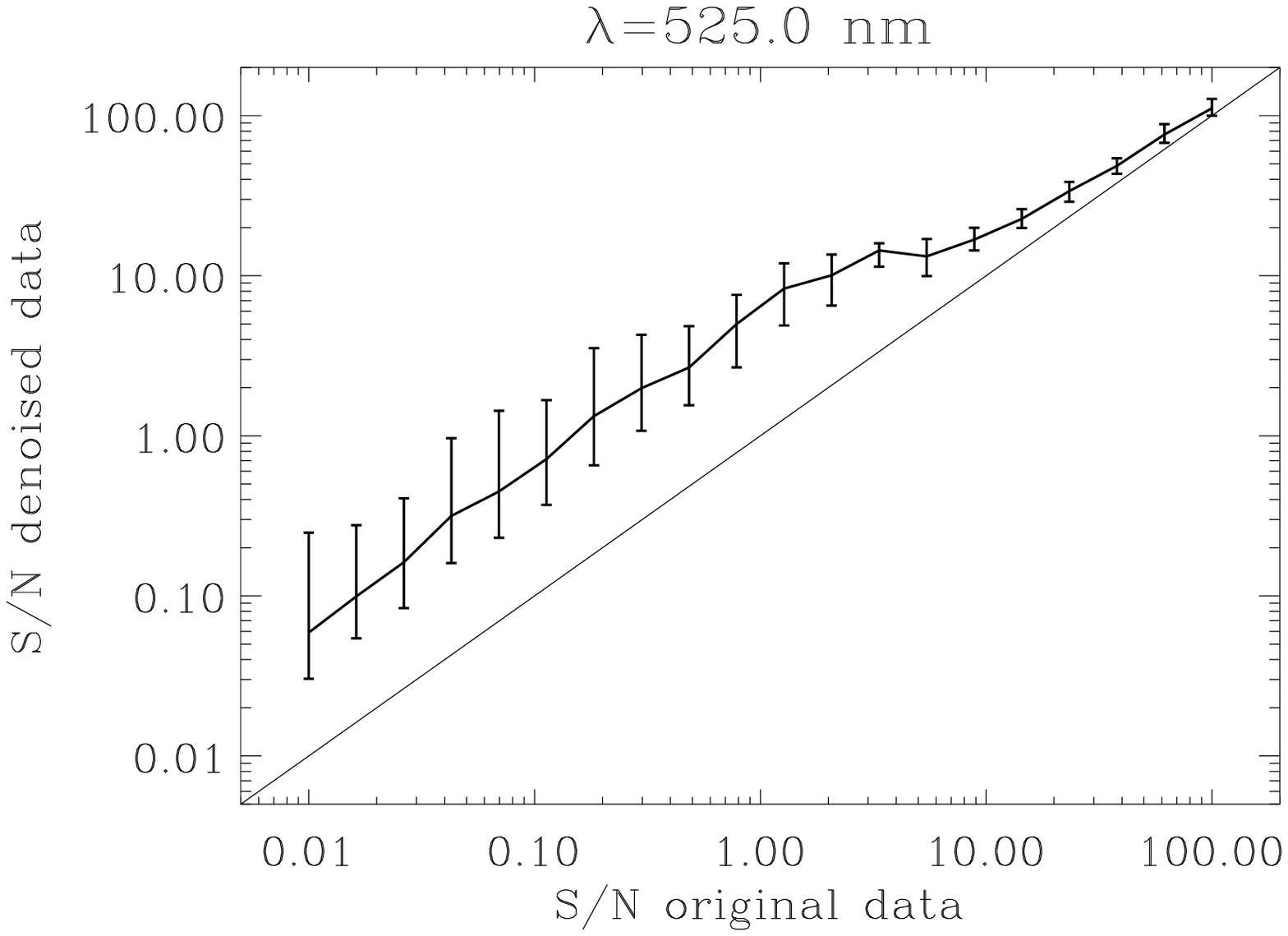}
\caption{Signal to noise ratio of the filtered data versus the original $S/N$ of
the observations for Stokes $Q$. The error bars indicate the 68\% confidence interval obtained in
a Montecarlo procedure with different noise realizations.
}
\label{finalQ}
\end{figure*}

The lowest the noise level the largest the
spectro-polarimetric information contained in the observations. Consequently, the $S/N$ of
the filtered data will be better if the $S/N$ of the observations is not
dramatically small. Of course, when the noise level in the observations is
negligible, the filtering procedure leads to a small improvement. Figure \ref{final}
presents the general trend in the improvement of the $S/N$ after the PCA denoising
is applied. The same applies to Fig. \ref{finalQ} for the case of Stokes $Q$.
The solid line in Fig. \ref{final} shows the $S/N$ of the filtered data versus the $S/N$ of
the observations. Again, we define the $S/N$ of the filtered data as the ratio between
the amplitude of the Stokes $V$ amplitude of the 630.2 nm line (10$^{-4}$ I$_\mathrm{c}$) and the standard deviation of the
difference between the filtered and the original profile. In order to estimate the statistical significance
of these values, we have estimated confidence intervals using a Montecarlo approach. The PCA denoising 
procedure has been applied to each line for 100 different realizations of each individual standard deviation of
the noise. The confidence intervals are obtained as the positions around the most probable value that
enclose 68\% of the probability.

Although the results are slightly
different for each spectral line, they share the same behavior. If the $S/N$ of
the observations is below $\sim 0.5$, the amount of polarimetric information that we can extract from the spectrum is very small. Therefore, the analysis based on principal components is of reduced
application. In this case, none of the eigenvalues of the cross-product matrix fulfill the
selection criterium and we use only the first eigenvector with detection purposes. However,
it is important to be cautious in this case, as already presented in Section \ref{sec:procedure}.
When the signal to noise is at least $0.6-0.7$, there is an improvement on the $S/N$ of the 
filtered signal. For instance, note that with a $S/N$ of 0.7 in the observations we increase 
it in almost one order of magnitude. Finally, as expected, if the noise level is very small we do not
improve the $S/N$ in the filtered data. Summarizing, as aparent from 
Figs. \ref{final} and \ref{finalQ}, the goal of having a $S/N \sim 1$ can be
accomplished with an observed spectrum with $S/N \sim 0.1$.

\section{Connection with previous approaches}
\subsection{Line addition}
The line addition technique \citep{semel96} consists of adding all the spectral lines
together for equal velocity displacements from line center. The photon noise and the 
blends are supposed to add in an incoherent way and the polarimetric information is supposed 
to add coherently. Consequently, one ends up with a mean profile with a considerably higher 
$S/N$, an increase that is roughly proportional to the square root of the number of added lines. It is a
very useful technique to detect polarisation in stellar spectra. The drawback is that the profile
is difficult to analyze. The mean profile is not a spectral line because its behavior with
the magnetic field is not the same as if it were a standard spectral line.

Under the PCA approach, it is also possible to retrieve the mean profile of the observations. 
The following expression, obtained after some simple algebra from Eqs. (\ref{ec2}),
gives the average profile $\vec{P}$:
\begin{equation}
\vec{P}=\frac{\sum_{jk} C_{jk} \vec{b}_k}{N_\mathrm{obs}},
\label{eq:mean_profile}
\end{equation}
where the index $k$ indicates the eigenvector and $j$ refers to the observation. The quantity
$C_{jk}$ is the projection of the observation $j$ onto the eigenvector $k$.
The previous analysis can also be done using only the first eigenvector. In this case, we end up
with the most common pattern in the data. Since we are using the cross-product matrix (and not the
covariance matrix in which the mean is substracted from the observations), the following 
average profile (reconstructed using only the first eigenvector) is very close to the mean 
profile of Eq. (\ref{eq:mean_profile}):
\begin{equation}
\vec{P}_1=\frac{\sum_{j} C_{j1} \vec{b}_1}{N_\mathrm{obs}}.
\end{equation}
The reason is that the first eigenvalue is the largest one and contains most of the 
variance of the data set. Then, as the eigenvalues drop rapidly, the rest of 
eigenvectors are much less important. The following analysis demonstrates that the PCA denoising can be made
equivalent to the line addition technique, being also a suitable technique to
detect magnetic signals in stars.

\subsection{Least-Squares Deconvolution}
The Least-Squares Deconvolution method presented
by \cite{donati97} is a variation of the line addition. 
It is based on the following two hypothesis: the lines are assumed
to be in the weak field regime of the Zeeman effect and there is a common pattern to all
spectral lines. Since the lines are assumed to be in the weak field regime of the Zeeman
effect, their Stokes $V$ profiles are proportional to the longitudinal component of the
magnetic field and the proportionality constant depends on the spectral line. This induces
that the resulting LSD Stokes $V$ profile is also linear in the longitudinal component of the
magnetic field making the analysis possible in terms of a pseudo-line with an effective
average Land\'e factor. The assumption that all the Stokes profiles of all the spectral
lines are proportional to a common ``mean Zeeman signature'' reduces the problem to the
following linear system of equations:
\begin{equation}
\vec{V}=\hat{W}\vec{Z},
\label{sistema_lineal}
\end{equation}
where $\vec{V}$ is a vector of length $N_\mathrm{obs} N_\lambda$ containing the
observed line profiles. The matrix $\hat{W}$, of size $N_\mathrm{obs} N_\lambda \times N_\lambda$, contains the weights $w$
for each spectral line which, in this approximation, are defined by means of its
Land\'e factor $g$ and its central depth $d$:
\begin{equation}
w=g\lambda d.
\label{eq:lsd_scaling}
\end{equation}
The vector $\vec{Z}$ is the so-called LSD profile, which is the characteristic Zeeman
signature of the star. One of the weakest points of this technique is that the weights
have to be proposed a priori. Once the weights are imposed, the linear system has the 
following least-squares solution:
\begin{equation}
\vec{Z}=(\hat{W}^t\hat{S}^2\hat{W})^{-1} \hat{W}^t\hat{S}^2\hat{V},
\end{equation}
where the matrix $\hat{S}$ is a diagonal matrix containing the inverse of the 
error bar of each spectral pixel. The previous solution is obtained using the weighted
pseudo-inverse of the matrix $\hat{W}$.

The examples shown in Figs. \ref{sn0113} and \ref{sn0785} are very close to what LSD
represents because we use only the first eigenvector for the reconstruction. By so doing,
we assume that all spectral lines can be reproduced by a common structure 
(the first eigenvector), the only difference between them being a scale 
factor (the projection of each observation along the first eigenvector). Instead of 
assuming the weight for each line, PCA naturally retrieves the common pattern in the 
observations and its intrinsic scale factor. However, note that the lowest the S/N the highest the 
dispersion of the PCA coefficients. Then, for 
extremely low S/N we must be careful on the interpretation of this coefficient in terms
of physical parameters.

\begin{figure}[!t]
\includegraphics[width=\columnwidth]{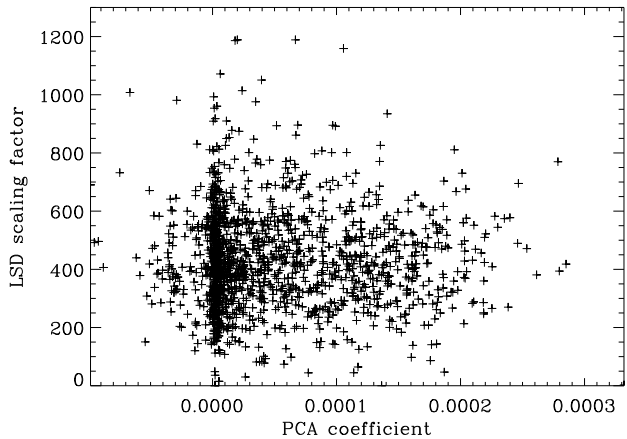}
\caption{Scatter plot showing the value of the projection of $\sim$2000 observed line profiles
onto the first eigenvector versus the LSD scaling factor associated with each line.}
\label{pca_vs_lsd}
\end{figure}

PCA can be understood as a generalization of the basic idea of LSD in the
sense that each spectral line is now a linear
combination of several particular functions. 
In order to investigate the relation between PCA and LSD, we present in Fig. \ref{pca_vs_lsd}
a scatter plot showing the value of the first PCA coefficient and the LSD scaling
factor, given by Eq. \ref{eq:lsd_scaling}. This plot has been obtained for $\sim$2000 spectral
lines without any noise added. Each spectral line can be contaminated with surrouding spectral lines in
the same spectral range, inducing negative projections on the first PCA eigenvector. Also note that
a reduced group of lines can have negative Land\'e factors, giving also negative projections
on the first PCA eigenvector. The values are unimportant since both the amplitudes of the LSD profile and 
the first eigenvector are not equivalent. However, the plot shows no apparent correlation between the two coefficients. In this 
particular case, even if the concepts of PCA and LSD are similar, the stellar Zeeman profile retrieved with LSD 
is not the common pattern in the data (which is clearly the first eigenvector). 
In a strict mathematical sense, PCA is not directly related to LSD but to the more general Total Least-Squares 
\citep[TLS;][]{huffel_tls02}. The standard linear least-squares method attributes all error to the dependent variables ($\vec{V}$ in our
case) and it minimizes the distance between the observations and the 
linear fit as measured along a particular axis direction. The weight matrix $\hat{W}$ is assumed to be
known without error. On the contrary, the linear TLS method allows errors in both the dependent and
independent variables ($\vec{V}$ and $\hat{W}$) and minimizes the \textit{perpendicular} distance to 
the linear fit. PCA is one of the methods that can be used to solve the linear TLS problem.

\section{Conclusions}

In this paper PCA is used to detect correlations 
between different velocity points and different spectral lines. The first principal component 
can be used to detect magnetic activity in stars. But the most important application of PCA is the 
denoising of individual spectral lines of stellar spectropolarimetric observations. The capabilities of the method are analysed using
numerical simulations. By assuming that the contaminating noise presents negligible correlation, 
we are able to isolate the signal from the noise. We have demonstrated that improvements close to one order of magnitude
in the signal to noise ratio per spectral line are typical for the present quality of
observed stellar polarised spectra. However, note that, although the method filters the noise in
each individual spectral line, the information contained in all of them is taken into account. 
The increase in the $S/N$ facilitates the future analysis of individual lines with 
standard techniques based on polarised radiative transfer theory.

The PCA denoising technique relies only on one free parameter: the number of PCA eigenvectors included 
in the reconstruction of the signal. We have suggested an automatic criterium for its selection
that works well on average. However, better denoising results can be obtained if one carefully
analyzes the resulting PCA eigenvectors and only selects those that carry an important amount of
signal as compared to the noise. 

Although the algebra is different, the PCA denoising technique is, in some way, related to other successful 
techniques for the detection of magnetic signals in stars as the line addition technique \citep{semel96} 
and the LSD procedure \citep{donati97}. Since we use the cross-product matrix the first principal component 
is very close to the mean profile obtained with the line addition. Moreover, PCA is directly related to 
Total Least-Squares, a method that can be seen as a generalization of LSD.

\begin{acknowledgements}
This research has been partially funded by the
Spanish Ministerio de Educaci\'on y Ciencia
through project AYA2007-63881.
\end{acknowledgements}


\end{document}